\documentclass[a4paper]{book}
\usepackage[utf8]{inputenc}
\usepackage{hyperref}
\usepackage{graphicx}
\usepackage{float}

\makeatletter

\makeatletter
\newcommand{\chapterauthor}[1]{%
  {\parindent0pt\vspace*{-25pt}%
  \linespread{1.1}\large\scshape#1%
  \par\nobreak\vspace*{35pt}}
  \@afterheading%
}
\makeatother

\usepackage{chapterbib}

\makeatother

\begin{document}


\chapter{Dissipative Kerr solitons in optical microresonators$^*$}

\chapterauthor{
Tobias Herr\footnotemark[1]$^,$\footnotemark[2]\\ 
Michael L. Gorodetsky\footnotemark[3]$^,$\footnotemark[4]\\
Tobias J. Kippenberg\footnotemark[1]
}

\renewcommand{\thefootnote}{\fnsymbol{footnote}}
\footnotetext[1]{To appear in {\it Nonlinear optical cavity dynamics}, ed. Ph. Grelu }
\renewcommand{\thefootnote}{\arabic{footnote}}
\footnotetext[1]{\'{E}cole Polytechnique F\'{e}d\'{e}rale de Lausanne (EPFL), CH-1015 Lausanne, Switzerland}
\footnotetext[2]{Centre Suisse d’\'{E}lectronique et de Microtechnique (CSEM),  Neuchâtel, Switzerland}
\footnotetext[3]{Faculty of Physics, M. V. Lomonosov Moscow State University, Moscow, Russia}
\footnotetext[4]{Russian Quantum Center, Skolkovo, Russia}

\bigskip


\textbf{Abstract:} This chapter describes the discovery and stable
generation of temporal dissipative Kerr solitons in continuous-wave (CW)
laser driven optical microresonators. The experimental signatures
as well as the temporal and spectral characteristics of this class
of bright solitons are discussed. Moreover, analytical and numerical
descriptions are presented that do not only reproduce qualitative
features but can also be used to accurately model and predict the
characteristics of experimental systems. Particular emphasis lies
on temporal dissipative Kerr solitons with regard to optical frequency
comb generation where they are of particular importance. Here, one
example is spectral broadening and self-referencing enabled by the
ultra-short pulsed nature of the solitons. Another example is 
dissipative Kerr soliton formation in integrated on-chip microresonators
where the emission of a dispersive wave allows for the direct generation
of unprecedentedly broadband and coherent soliton spectra with smooth
spectral envelope.

\section{Introduction to optical microresonator\\ Kerr-frequency combs \label{sec:Introduction}}

Ultra high quality-factor (Q) optical whispering-gallery mode resonators
have long been considered promising for their nonlinear optical properties
since their discovery by Braginsky and co-workers in 1989 \cite{Braginsky1989}.
Early works included the observation of bistability, thermal nonlinearity\cite{Ilchenko1992a},
Kerr-nonlinearity at cryogenic temperature\cite{Treussart1998}, stimulated
Raman- \cite{Spillane2002} and Brillouin-scattering\cite{Grudinin2009a},
electro-optic effect\cite{Cohen2001,Ilchenko2003b}, as well as second\cite{Ilchenko2004}
and third\cite{Carmon2007c} harmonic generation. In particular, the
third order Kerr-nonlinearity can give rise to parametric oscillations
in an optical microresonator. In this four-photon process two pump
photons are converted to one signal and one idler photon (i. e. degenerate
four-wave mixing) where the photonic energy is conserved. These parametric
oscillations (sometimes called hyper-parametric to distinguish them
from three-photon oscillations associated with the second order optical
nonlinearity), despite being well-known in nonlinear optics for decades\cite{Klyshko1988},
were observed in ultra high-Q toroidal fused silica and crystalline
resonators only in 2004\cite{Kippenberg2004a,Savchenkov2004c}. These
studies demonstrated the dramatic reduction of threshold power for
nonlinear optical oscillation owing to the $1/Q^{2}$ scaling of the
threshold\cite{Braginsky1989}.

\begin{figure}[h]
\includegraphics[width=1.0\columnwidth]{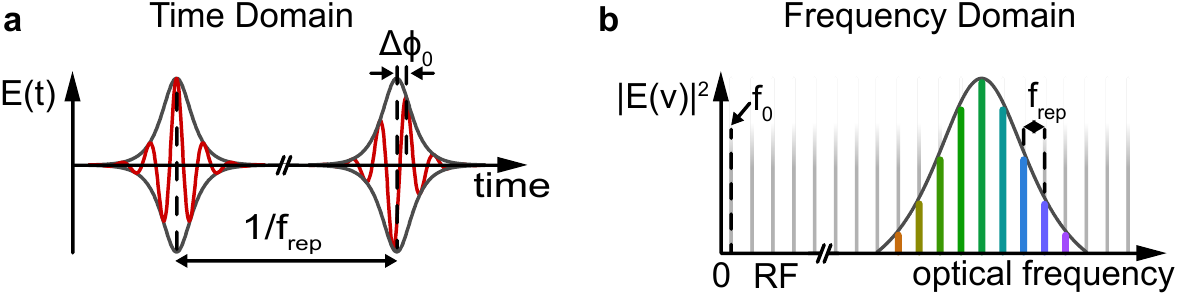}

\protect\protect\protect\protect\caption{\label{fig:CombsGeneral}Time and frequency domain picture of mode-locked
laser based frequency combs. A periodic train of pulses with a pulse
repetition rate $f_{\mathrm{rep}}$ (panel a) corresponds to a comb
spectrum of equidistant lines in the frequency domain (panel b). The
line spacing is given by $f_{\mathrm{rep}}$. The offset $f_{\mathrm{0}}$
of the frequency comb spectrum relates to the carrier-envelope phase
shift $\Delta\phi_{0}$ between two consecutive pulses via $f_{0}=f_{\mathrm{rep}}\cdot\Delta\phi_{0}/(2\pi)$.
The two parameters $f_{\mathrm{rep}}$ and $f_{\mathrm{0}}$ fully
define all comb frequencies $f_{n}=n\cdot f_{\mathrm{rep}}+f_{0}$.}
\end{figure}

In 2007 it was discovered\cite{DelHaye2007} that Kerr-nonlinear optical
microresonators can give rise to optical frequency comb\cite{Udem2002,Cundiff2003}
generation. The surprising cascade and mode proliferation associated
with four-wave mixing (FWM) was demonstrated to lead to a broadband
series of optical lines equidistantly spaced in the optical frequency
domain, that is an optical frequency comb. The discovery of these
so-called microresonator \textit{Kerr-combs} established optical microresonators
as tools for precision frequency metrology, a connection not made
before. Such frequency comb spectra are conventionally derived from
mode-locked femto-second lasers. In this way, Kerr-combs broke with
the conventional dogma in the frequency metrology community, that
an optical frequency comb requires a mode-locked pulsed laser source.
Different from conventional mode-locked laser based combs, regular
microresonator Kerr-combs are not pulsed in the time domain. By comparing
the comb spectrum from a microresonator with a conventional mode-locked
fiber laser comb, the equidistance of the comb lines could be proven
at the level of 1 part in $10^{17}$. Since this work the field of
microresonator Kerr-combs has increased substantially. In the subsequent
years, Kerr-combs have been demonstrated in a variety of platforms,
including crystalline resonators\cite{Savchenkov2008}, CMOS compatible
platforms such as silicon nitride ($\mathrm{Si_{3}N_{4}}$)\cite{Foster2011a,Levy2010,Razzari2010},
Hydex glass\cite{Moss2013}, as well as aluminum nitride \cite{Jung2013}
or diamond\cite{Hausmann2014}. In addition to allowing for miniaturization
and chip-scale integration of frequency comb oscillators, microresonator
based Kerr-combs enable to attain wide comb line spacings in the technologically
relevant $10-100$ $\mathrm{GHz}$ range\cite{Del'Haye2008,Johnson2012},
which is not easily accessible using mode-locked laser frequency combs.
Such large mode spacing is of particular interest in coherent telecommunication\cite{Pfeifle2014},
spectrometer calibration for astronomy\cite{Steinmetz2008,Murphy2007},
low noise microwave\cite{Savchenkov2008a,Li2012} and arbitrary optical
waveform generation\cite{Jiang2007}. Optical temporal dissipative
cavity Kerr solitons, predicted theoretically\cite{Barashenkov1996,Wabnitz1993}
and first realized experimentally in optical fibers\cite{Leo2010},
have played a decisive role in allowing to overcome noise processes
associated with comb formation in microresonators. As detailed later,
temporal \textit{dissipative Kerr solitons} (DKS) provide a deterministic way to generate
broadband, coherent, and spectrally smooth optical frequency combs,
which are in addition amenable to an analytical theory and numerical
simulation. These developments, along with the experimental generation
of single soliton states in crystalline resonators\cite{Herr2013,Herr2014}
and integrated $\mathrm{Si_{3}N_{4}}$ resonators\cite{Brasch2014b},
make microresonators suitable for precision frequency metrology and
many related applications. Owing to the pulsed nature of the solitons
(in contrast to the earlier Kerr-combs), microresonators can now provide
equivalent counterparts to mode-locked laser regarding the time domain
properties. This allows using methods of spectral transfer and broadening
as well as self-referencing techniques \cite{Reichert1999,Telle1999,Ramond2002}
that have been developed for mode locked laser-systems. In this way
DKS in microresonators provide a route to
synthesize absolute optical frequencies from an RF or microwave signal
and the opportunity to use microresonators for counting the cycles
of light.

\section{Resonator platforms}

Parametric frequency conversion in Kerr-nonlinear microresonators,
first observed in toroidal fused silica and crystalline microresonators,
has now been observed in a wide variety of geometries and platforms,
including microspheres\cite{Agha2007,Agha2009} and integrated, planar
ring-type resonators, such as those based on $\mathrm{Si_{3}N_{4}}$\cite{Turner2008}.
The work presented in this chapter on Kerr soliton generation
has been carried out with crystalline magnesium fluoride $\mathrm{MgF_{2}}$
resonators and $\mathrm{Si_{3}N_{4}}$ integrated microrings, that
are briefly described in the next sections.

\begin{figure}[tbh]
\includegraphics[width=1.0\columnwidth]{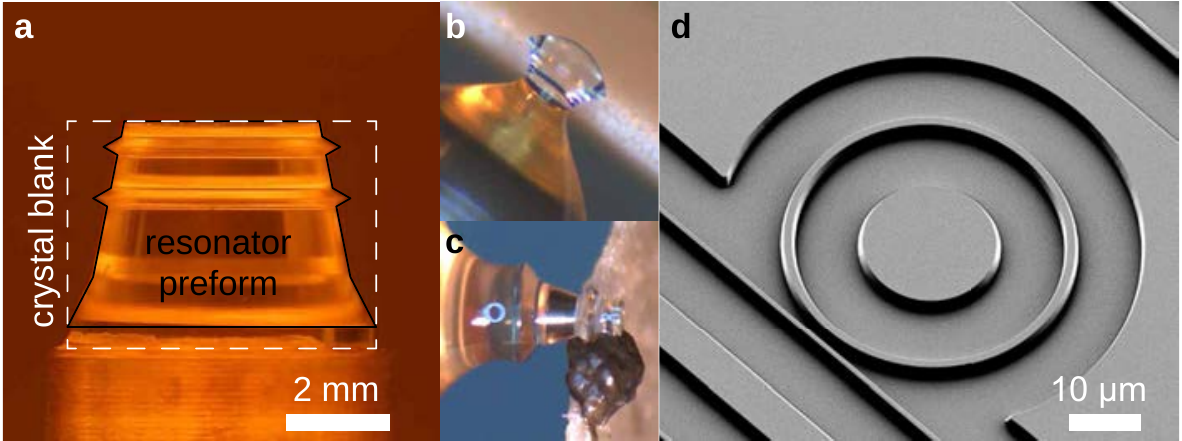}
\protect\protect\protect\protect\caption{Microresonator platforms. (a) Diamond
turned magnesium fluoride resonator containing two protrusions that
confine high-Q optical whispering-gallery modes. (b) Manually shaped
and polished resonator. (c) Polishing of a diamond turned preform.
(d) Scanning electron micrograph of a silicon nitride microresonator
before applying the fused silica cladding.\label{fig:Microresonators}}
\end{figure}

\subsection{Ultra high-Q MgF$_2$ Crystalline microresonators }

The first microresonator platform in which temporal DKS have been observed 
experimentally\cite{Herr2013} were crystalline
resonators made from magnesium fluoride$(\mathrm{MgF_{2}})$. Crystalline
optical resonators were first introduced in \cite{Ilchenko2000} and
further development demonstrated that polishing of crystalline materials
can lead to exceptionally high Q-factors (exceeding $10^{11}$ \cite{Savchenkov2004,Savchenkov2007b}).
This method therefore extended the ultra high-Q attained with surface
reflow methods\cite{Braginsky1989,Armani2003} to crystalline materials.
Figure \ref{fig:Microresonators} shows an ultra high-Q crystalline
resonator made from a $\mathrm{MgF_{2}}$ cylindrical preform, which
is prestructured with protrusions using precision diamond turning.
After prestructuring a series of polishing steps using diamond films
and slurries are used to attain a low roughness surface finish and
ultra high-Q optical modes. To excite the whispering-gallery modes
evanescent optical coupling via a tapered optical fiber can be used\cite{Knight1997,Spillane2003}.
This is possible as the refractive index of $\mathrm{MgF_{2}}$ is
lower than that of fused silica fibers. The use of thin tapered fiber
for coupling is feasible despite the large volume of the employed
(typically mm scale) resonators, due to the exceptionally high-Q,
which reduces the necessary coupling rates required for critical coupling.
Typically the resonance width is in the range from $50-500$ kHz corresponding
to Q values of $10^{8}$ to several $10^{9}$. The nonlinear frequency
conversion threshold is reached at optical powers below $1$ mW. For
soliton generation pump powers of the order of $10$ mW are typical.

\subsection{Integrated photonic chip microring resonators}

From a future application perspective microresonator platforms amenable to wafer-scale processing can be advantageous. A key
challenge in this context is attaining sufficiently high-Q for efficient
nonlinear parametric frequency conversion. One platform that is particularly
well suited is based on silicon nitride ($\mathrm{Si_{3}N_{4}}$)
waveguides embedded in fused silica. is a material already used in
the micro-electronic industry (part of the CMOS process) and is well
suited for integrated photonic waveguides. Its refractive index of
$n\approx2$ enables tight optical confinement waveguides and microring
resonators and the high bandgap ($\sim3$ eV) mitigates two (or higher
order) photon absorption in the telecommunication band. While the
achieved Q-factors ($\mathcal{\sim}10^{6}$) in integrated devices
are still many orders of magnitude below that of polished crystalline
resonators , higher Kerr-nonlinearity and tighter optical confinement,
lead to high effective nonlinearities that are almost three orders
of magnitude larger than in crystalline resonators. This has enabled
parametric oscillations to be accessible with pump powers of $\mathcal{\sim}100\mathrm{mW}$
as demonstrated first in Refs. \cite{Levy2010,Levy2009}. Subsequent
work demonstrated low phase noise comb operation\cite{Herr2012} via
nonlinear synchronization, as detailed later. In addition $\mathrm{Si_{3}N_{4}}$
microresonators were the first microresonator platform that allowed
for the observation of soliton induced Cherenkov radiation\cite{Akhmediev1995}
(or dispersive wave emission), i.e. the dynamics of solitons in the
presence of higher order dispersion\cite{Brasch2014}. Fabrication
of the resonators proceeds by lithography, etching and a final encapsulation
technique in which the waveguides are clad with fused silica. The
image of a final (but not yet clad resonator) is shown in Figure \ref{fig:Microresonators}.
Coupling to and from the chip is achieved with inverse tapered waveguides,
adiabatically converting the mode spot diameter from a tight nanophotonic
waveguide mode to a large area mode that can be excited with a lensed
single-mode optical fiber.

\section{Physics of the Kerr-comb formation process\label{sec:PhysicsCombFormation}}

\begin{figure}[tbh]
\includegraphics[width=1.0\columnwidth]{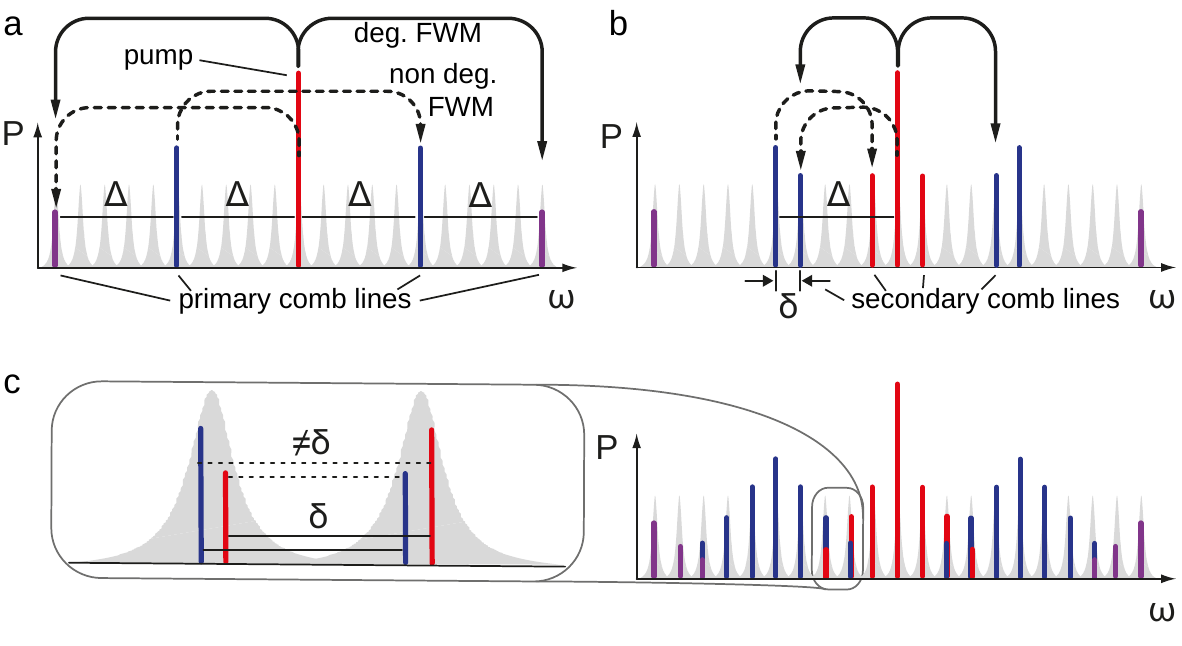}

\protect\protect\protect\protect\caption{\label{fig:Subcomb-formation-process}Universal Kerr-comb formation
processes. (a) Formation of primary sidebands (b) Formation of subcombs
(c) Overlap between inconsistent subcombs can lead to multiple lines
per cavity resonance and explains noise phenomena in Kerr-combs.}
\end{figure}

\begin{figure}[tbh]
\includegraphics[width=1.0\columnwidth]{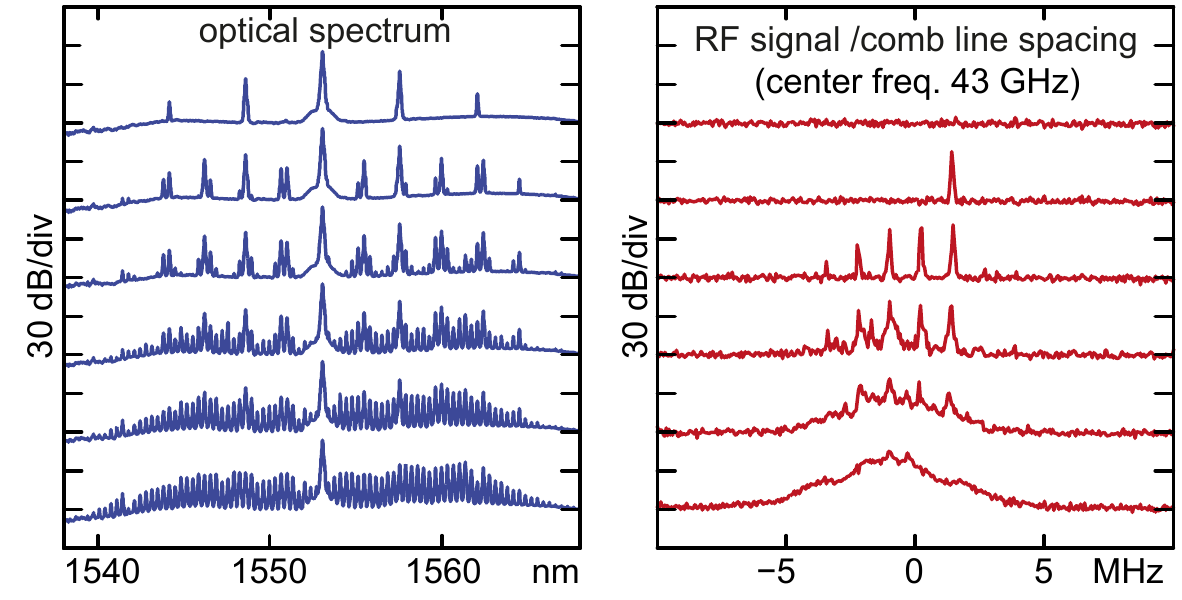}

\protect\protect\protect\protect\caption{\label{fig:Emergence-of-Noise}Kerr-comb formation and noise. (a)
Formation of the optical spectrum as the laser is tuned into resonance
and the intracavity power increases. (b) Comb-line spacing measured
as the radio-frequency (RF) beatnote between neighboring comb lines.
Multiple and broad beatnotes indicate multiple and inconsistent line
spacings present in the comb spectrum.}
\end{figure}

In a simplified picture, the Kerr-comb formation in microresonators
starts with an initial degenerate FWM process, generating a pair of
sidebands symmetrically spaced in frequency around the pump laser
in the resonances adjacent to the pumped resonance. In this process
the free spectral range (FSR) defines the frequency spacing of the
sidebands from the pump laser. A subsequent cascade of non-degenerate
four-wave mixing processes generates a large series of sidebands in
frequency steps corresponding to the initially defined line spacing.
The bandwidth of the comb is limited by the dispersion of the resonator
that causes a mismatch between the optical sidebands and the resonance
frequencies. This mismatch increases with the spectral distance to
the pump wavelength. While it explains well the early results of Kerr-frequency
comb generation with typically THz-line spacing and moderate bandwidth,
this picture had to be refined when attempting broader optical bandwidth
and narrower line spacing (below $50$ GHz) in toroidal \cite{Del'Haye2011},
crystalline \cite{Savchenkov2008} and integrated $\mathrm{Si_{3}N_{4}}$
resonators\cite{Foster2011a,Ferdous2011,Okawachi2011}. Here unexpected
noise phenomena were observed in the form of broad optical lines,
high amplitude noise and loss of coherence. These phenomena appeared
to be independent of the chosen microresonator platform. Indeed, the
origin of this noise can be explained by the universal properties
of the comb formation mechanism\cite{Herr2012}. While in early microresonator
comb experiments, due to the wide line spacing, the mode proliferation
occurred indeed on adjacent resonators modes, the latter is typically
not the case in resonators with narrow FSR. Here the first pair of
primary sidebands generated by the initial FWM process can occur in
resonances that are widely separated from the pump (by multiple FSR).
The process is depicted schematically in Figure \ref{fig:Subcomb-formation-process}.
Using the nonlinear coupled mode equations, as introduced in the next
section, one can show that the mode spacing between the primary modes
is essentially defined by the (normalized) ratio of second order dispersion
parameter and resonance width. If the resonator has a comparatively
narrow FSR, then the initial sidebands are separated by a frequency
interval $\Delta$ that encompasses a high number of FSR intervals.
Further four-wave mixing can fill up the unpopulated resonances by
formation of secondary sidebands separated by a smaller frequency
spacing $\delta$. In other words, sub combs with $\delta$-line spacing
form around the $\Delta$-spaced primary lines. As, however, the primary
spacing $\Delta$ is not guaranteed to be an integer multiple of the
secondary spacing $\delta$, the generated lines can not generally
form a consistent comb, which manifests itself in the previously mentioned
noise. This hypothesis of comb formation has been experimentally tested
in \cite{Herr2012} using a fiber laser frequency comb as a reference
to reconstruct the frequency components of the Kerr-comb. Another
key finding of this work is that in contrast to the early model of
comb formation, more than one single comb line can occupy a given
resonance (cf. Figure \ref{fig:Subcomb-formation-process} and Figure
\ref{fig:Emergence-of-Noise}).

\begin{figure}[tbh]
\includegraphics[width=1.0\columnwidth]{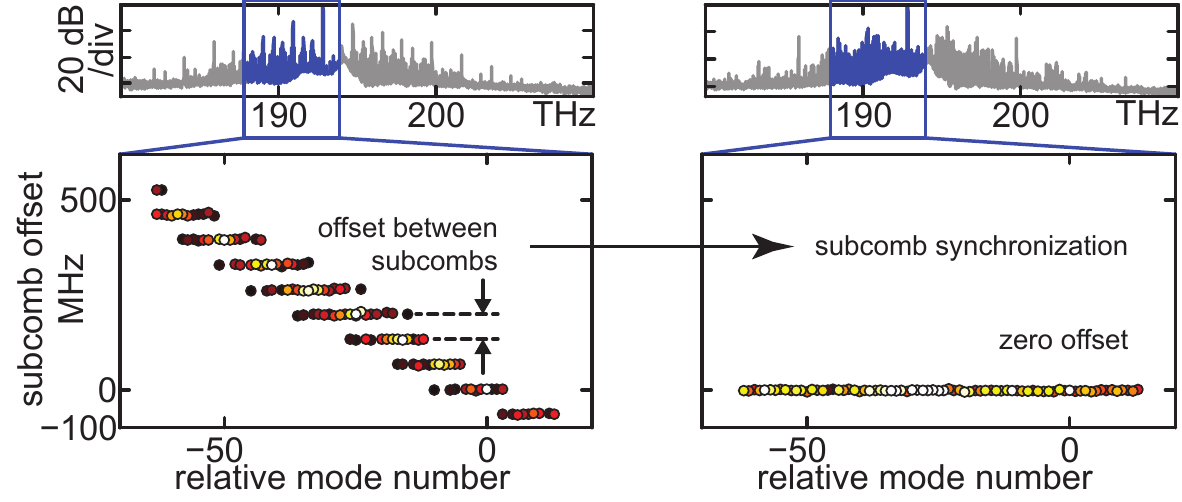}

\protect\protect\protect\protect\caption{\label{fig:Subcomb-synchronization}Subcomb synchronization in a $\mathrm{Si_{3}N_{4}}$
optical microresonator achieved by changing the pump laser wavelength..}
\end{figure}

Despite these noise processes, regimes have been found where broadband
coherent Kerr-combs can be generated. The first observed transition
to such a regime of low phase noise was presented in \cite{Herr2012}.
Here the subcombs defined by $\Delta$ and $\delta$ can be synchronized
by changing the pump laser frequency and power so that $\Delta$ and
$\delta$ become commensurate. Once close enough to the commensurate
state, nonlinear synchronization and locking sets in and creates a
phase coherent optical frequency comb\cite{Herr2012} (albeit not
pulsed in the time domain). Such synchronization has been experimentally
observed and studied by reconstruction of the frequencies of the Kerr-comb
lines as shown in Figure\ref{fig:Subcomb-synchronization}. Once the
Kerr-comb falls into such a synchronized state it remains stable against
fluctuations in the pump laser parameters and can be used for further
experiments.

Later experiments have revealed that the nonlinear locking process
can be captured by the Adler equation, that is known from injection
locking of lasers\cite{DelHaye2014}. Kerr combs driven into a low-phase
noise, coherent state have been used in applications, such as coherent
communications\cite{Pfeifle2014}, where they are used to provide
a multitude of data channels from a single laser, enabling terabit
per second data-communication. Yet, despite the promising prospects
a central missing element has been ways to reliably achieve such low
phase noise combs, with smooth spectral envelope. A surprising discovery
has been made in 2012\cite{Herr2012b,Herr2013}, when it was demonstrated
that the generated subcombs can not only undergo nonlinear synchronization,
but moreover can seed the formation of temporal DKS in optical microresonators. 
This discovery provides not only an opportunity
to study DKS in microresonators, but moreover
a powerful tool to generate low phase noise combs with broad spectral
bandwidth that can be used in a variety of applications. Before turning
to soliton formation, we review first the basic formalism describing
Kerr-nonlinearity driven parametric oscillations.

\subsection{Nonlinear coupled mode equations \label{sec:NonlinearCoupledMode Equations}}

The theoretical analysis of parametric oscillations, frequency comb
and soliton formation is a complex nonlinear mathematical problem
for which several approaches have been proposed \cite{Kippenberg2004a,Matsko2005a,Agha2009,Maleki2010,Matsko2011a,Chembo2010d}.
Besides description in the time domain the modal expansion approach
in the frequency domain \cite{Chembo2010,Chembo2010b} has proven
particularly useful in the context of microresonators as it allows
easily to take into account particularities in the mode structure.
The system of equations describing the dynamics of each optical mode
$A_{m}$ in the slowly varying envelope approximation can be obtained
from the nonlinear wave equation in conjunction with the quantum Langevin
equations: 
\begin{eqnarray}
\frac{\partial A_{m}}{\partial t} & = & -\frac{\kappa_{m}}{2}A_{m}+\delta_{m,m_{0}}\sqrt{\frac{\kappa_{\mathrm{ex,m}}P_{in}}{\hbar\omega_{0}}}e^{-i(\omega_{p}-\omega_{0})t}\nonumber \\
 & + & ig\cdot\sum_{m',m'',m'''}\!\!\!\!\!\!\!\!\Lambda_{m}^{m'm''m'''}{\cal D}_{m}^{m'm''}A_{m'}A_{m''}A_{m'''}^{*}e^{-i(\omega'_{m}+\omega''_{m}-\omega'''_{m}-\omega_{m})t}\nonumber \\
 & + & \sqrt{\kappa_{\mathrm{i},m}}\delta s_{\mathrm{i},m}(t)+\sqrt{\kappa_{\mathrm{ex},m}}\delta s_{\mathrm{ex},m}(t).
\end{eqnarray}
Here $\kappa_{m}$ denotes the cavity decay rate of mode $m$ with
eigenfrequency $\omega_{m}$, $\hbar$ is the reduced Planck constant.
The cavity decay rate is composed of intrinsic ($\kappa_{\mathrm{i},m}$) and
external coupling ($\kappa_{\mathrm{ex},m}$) loss rates such that the total
loss rate is $\kappa_{m}=\kappa_{\mathrm{i},m}+\kappa_{\mathrm{ex},m}=\frac{\omega_{m}}{Q_{m}}$,
where $Q_{m}$ is the loaded quality-factor of mode $m$. One mode
($m_{0}$, $\omega_{0}$) is driven by a classical field with optical
power $P_{in}$ and frequency $\omega_{p}$. The terms $\delta s_{\mathrm{i},m}$
and $\delta s_{\mathrm{ex},m}$ designate quantum Langevin input noises with
commutation relations $[\delta s_{j}(t),\delta s_{j}^{\dagger}(t')]=\delta(t-t')$
and with zero mean ($\langle\delta s_{j}(t)\rangle=0$). The summation
over the nonlinear interaction terms is made over all azimuthal indices
for the nearly equidistant set of fundamental whispering-gallery eigenmodes,
satisfying the following conservation relation: $m'+m''-m'''=m$,
where ${\cal D}_{m}^{m'm''}$ is the degeneracy factor: ${\cal D}_{m}^{m'\neq m''}\!\!=\!2$
in case of cross-phase modulation and non-degenerate four-wave mixing,
and ${\cal D}^{m'm'}\!\!=1$ for self-phase modulation and hyperparametric
generation. We neglect processes where one of the frequencies is significantly
smaller ($\omega_{m}=\omega_{m'}-\omega_{m''}-\omega_{m'''}$) or
larger than other frequencies ($\omega_{m}=\omega_{m'}+\omega_{m''}+\omega_{m'''}$)
as it is the case for third harmonic generation (when in the degenerate
case ${\cal D}_{3m'}^{m'm'}=1/3$). The intermodal coupling factor
is given by 
\begin{eqnarray}
\Lambda_{m}^{m'm''m'''}=\frac{\int{\bf e}_{m'}{\bf e}_{m''}{\bf e}_{m'''}^{*}{\bf e}_{m}^{*}dV}{\int|{\bf e}_{0}|^{4}dV}\frac{\chi^{(3)}(\omega_{m}=\omega_{m'}+\omega_{m''}-\omega_{m'''})}{\chi^{(3)}(\omega_{0}=\omega_{0}+\omega_{0}-\omega_{0})},
\end{eqnarray}
where ${\bf e}_{m}$ is vector electric field distribution of the
mode $m$ and $\chi^{(3)}$ describes the third order Kerr-nonlinearity.
This factor is analytically different from that obtained in \cite{Chembo2010d,Chembo2010,Chembo2010b}
\footnote{The authors of \cite{Chembo2010d,Chembo2010,Chembo2010b} overlooked
in the derivation that the nonlinear polarization is time dependent
and can not be simply taken out from the time derivative in wave equation
\cite{Boyd2007}. This omission leads to unphysical asymmetry in $\Lambda$
over mode indices and loss of precise degeneracies important for analytical
considerations, but usually plays negligible role in numerical simulations
of narrow combs when $\Lambda\simeq1$.%
}. The effective nonlinear volume of the modes is given by: 
\begin{eqnarray}
V_{m}=\frac{(\int|{\bf e}_{m}|^{2}dV)^{2}}{\int|{\bf e}_{m}|^{4}dV}.
\end{eqnarray}
The latter may be found from the cavity eigenmodes' field distribution.
The modes are normalized such that $|A_{m}|^{2}$ corresponds to the
number of photons in a given cavity mode. The nonlinear coupling constant
$g$ (that is the Kerr frequency shift per photon) is defined as in
\cite{Matsko2005a}: 
\begin{eqnarray}
g & =\frac{3\hbar\omega_{0}^{2}\chi^{(3)}(\omega_{0})}{4\epsilon_{0}n^{4}(\omega_{0})V_{0}}=\frac{\hbar\omega_{0}^{2}cn_{2}(\omega_{0})}{n^{2}(\omega_{0})V_{0}}.
\end{eqnarray}
Here $\epsilon_{0}$ denotes the dielectric constant and $n$ and
$n_{2}$ are the refractive nonlinear optical indices. A seeding via
input noises is important for numerical simulations to initiate the
comb, however, it is not required in further analytical considerations
as their means vanish.

For the calculation of the field distribution and eigenfrequencies
in crystalline resonators we do not use the modes of a sphere as proposed
in \cite{Chembo2010,Chembo2010b} but asymptotic solutions obtained
for a spheroid \cite{Gorodetsky2007,Gorodetsky2011,Demchenko2013},
producing better approximations for the wide range of geometries,
or eigenfrequencies obtained via \texttt{COMSOL} finite element modeling. From
these approximations as well as from Sellmeier's equation for the
resonator material one can determine $V_{m}$ and eigenfrequencies
$\omega_{m}$ to find second and higher order dispersions. The latter
allows can be approximated by  
\begin{eqnarray}
\omega_{\mu}\approx\omega_{0}+D_{1}\mu+\frac{1}{2!}D_{2}\mu^{2}+\frac{1}{3!}D_{3}\mu^{3},\label{dispfreq}
\end{eqnarray}
where $\mu=m-m_{0}$. The parameter $D_{2}$ is related to the second
order group velocity dispersion (GVD) $\beta_{2}$ ($\beta_{j}=\frac{\partial^{j}\beta}{\partial\omega^{j}}$)
via $D_{2}=-\frac{c}{n}D_{1}^{2}\beta_{2}$, and $D_{3}$ relates
to the third order dispersion via $D_{3}=-\frac{c}{n}D_{1}^{3}\beta_{3}+3\frac{c^{2}}{n^{2}}D_{1}^{3}\beta_{2}^{2}\approx-\frac{c}{n}D_{1}^{3}\beta_{3}$.

It is convenient to remove the explicit time dependence in the nonlinear
equations with the substitution $A_{\mu}=\tilde{A}_{\mu}e^{i(\omega_{\mu}-\omega_{p}-\mu D_{1})t}$,
where $D_{1}/(2\pi)$is the mode spacing of the comb at the pump wavelength.
Moreover, the pump term is written as $F=\sqrt{\frac{\kappa_{c}P_{in}}{\hbar\omega_{0}}}$.
In this new basis the oscillations of each optical mode are considered
not around eigenfrequencies but around nearest frequencies on an equidistant
$D_{1}$ spaced-grid: 
\begin{eqnarray}
\frac{\partial\tilde{A}_{\mu}}{\partial t}= & - & [i(\omega_{\mu}-D_{1}\mu-\omega_{p})+\frac{\kappa_{\mu}}{2}]\tilde{A}_{\mu}+\delta_{\mu0}F\nonumber \\
 & + & ig\sum_{\mu',\mu'',\mu'''}\Lambda_{\mu}^{\mu'\mu''\mu'''}\tilde{A}_{\mu'}\tilde{A}_{\mu''}\tilde{A}_{\mu'+\mu''-\mu}^{*}.
\end{eqnarray}
If $L$ pairs of sidebands and the pump are considered, then the total
number of nonlinear terms in all $2L+1$ equations is $\frac{1}{3}(L+1)(8L^{2}+7L+3)$.

It is possible to get some useful conservation laws for this set of
equations. Calculating the evolution of the total number of photons
in the comb $\frac{dN}{dt}=\sum_{\mu=-L}^{L}\frac{dN_{\mu}}{dt}$,
with $\frac{dN_{\mu}}{dt}=\frac{dA_{\mu}}{dt}A_{\mu}^{*}+\frac{dA_{\mu}^{*}}{dt}A_{\mu}$
and assuming symmetry: $\Lambda_{\mu}^{\mu'\mu''\mu'''}+\Lambda_{\mu'''}^{\mu'\mu''\mu'''}=\Lambda_{\mu'}^{\mu\mu'''\mu''}+\Lambda_{\mu''}^{\mu\mu'''\mu'}$,
we find that most of the terms in the sum cancel each other: 
\begin{eqnarray}
\frac{dN}{dt}+\sum_{\mu=-L}^{L}\kappa_{\mu}N_{\mu}=(FA_{0}^{*}+F^{*}A_{0}),
\end{eqnarray}
which in case of $\kappa_{\mu}=\kappa$ simplifies to 
\begin{eqnarray}
\frac{dN}{dt}+\kappa N=2|F||A_{0}|\cos\phi_{0},\label{FirstIntegral}
\end{eqnarray}
where $\phi_{0}$ is the phase difference between $F$ and $A_{0}$.
This master equation for the whole optical spectrum may be useful
for example for the analysis of the  steady states. 

In a similar way one may find: 
\begin{eqnarray}
\sum_{\mu=-L}^{L}\mu\left(\frac{dN_{+\mu}}{dt}-\frac{dN_{-\mu}}{dt}\right)=-\sum_{\mu=-L}^{L}\mu\left(\kappa_{+\mu}N_{+\mu}-\kappa_{-\mu}N_{-\mu}\right),
\end{eqnarray}
which in the case of steady state means that the 'center of mass'
(i.e. the frequency center of photonic energy) of the spectrum is
conserved and coincides with the pump frequency as each four-wave
mixing process follows this conservation law.

In the following analysis we use further simplifications, assuming
equal parameters for all comb lines ($\kappa_{m}=\kappa$, $\Lambda_{m}^{m'm''m'''}=1$)
and renumbering comb lines with index ${\mu}$, starting from the
central driven mode $m_{0}$, i.e. $\mu=m-m_{0}$. In this case
it is possible to write the equations in a dimensionless way, as suggested
in \cite{Matsko2005a}: %
\begin{eqnarray}
a_{\mu} & = & \tilde{A}_{\mu}\sqrt{2g/\kappa},\nonumber \\
f & = & \sqrt{8g/\kappa^{3}}|F_{p}|,\nonumber \\
\zeta_{\mu} & = & 2(\omega_{\mu}-\mu D_{1}-\omega_{p})/\kappa\approx\zeta_{0}+D_{2}\mu{}^{2}/\kappa,\nonumber \\
\tau & = & \kappa t/2,
\end{eqnarray}
\begin{equation}
\frac{\partial a_{\mu}}{\partial\tau}=-[1+i\zeta_{\mu}]a_{\mu}+i\sum_{\mu^{\prime},\mu^{\prime\prime}}a_{\mu^{\prime}}a_{\mu^{\prime\prime}}a_{\mu^{\prime}+\mu^{\prime\prime}-\mu}^{*}+\delta_{0\mu}f.\label{coupled_modes_eq}
\end{equation}
In this form all frequencies, detunings and magnitudes are measured
in units of cavity resonance width so that $|a_{\mu}|^{2}=1$ corresponds
to the nonlinear mode pulling of one cavity resonance width (thresholds
for both single mode bistability and degenerate oscillations).

\subsection{Degenerate hyperparametric oscillations}

An analytic solution of the complete nonlinear system with many comb
lines in general case is hardly possible. It is possible, however,
to resolve equations for an arbitrary set of modes if only degenerate
processes caused by the pump as well as nonlinear mode pulling effects
corresponding to cross- and self-phase modulation are taken into account.
These effects are dominating as long as the energy in each of the
optical lines is small compared to the intracavity pump power and
when due to strong second order dispersion the resonator modes are
far enough from equidistance lines to suppress non-degenerate four-wave
mixing.

The following set of reduced equations qualitatively describes the
processes of initial comb formation, including widely spaced  primary
lines, switching between initially excited modes, chaotic behavior
and noisy RF signals with many frequencies: 
\begin{eqnarray}
\frac{\partial a_{0}}{\partial\tau}= & -[1+i\zeta_{0}]a_{0}+i(2\tilde{N}-|a_{0}|^{2})a_{0}+2i\sum_{\mu=1}^{L}a_{\mu}a_{-\mu}a_{0}^{*}+f,\nonumber \\
\frac{\partial a_{\mu}}{\partial\tau}= & -[1+i\zeta_{\mu}]a_{\mu}+i(2\tilde{N}-|a_{\mu}|^{2})a_{\mu}+ia_{-\mu}^{*}a_{0}^{2},\label{simplifiedset}
\end{eqnarray}
where according to the normalization $\tilde{N}=2gN/\kappa$. The
reduced model was compared with the full numeric model integrated
using the Runge-Kutta scheme. It shows good agreement concerning the
initially generated sidebands and qualitative agreement of the comb
in the noisy, chaotic regime.

With the substitution $a_{\mu}=\alpha_{\mu}e^{-i\phi_{\mu}},$ $\psi_{\mu}=2\phi_{0}-\phi_{\mu}-\phi_{-\mu}$
it is possible to search for steady state solutions with $\dot{\alpha}_{\mu}=0$
and $\dot{\phi}_{\mu}\equiv\delta\tilde{\omega}_{\mu}=\mathrm{const}$.
One may obtain a a steady state from the resulting equations when
$\dot{\phi}_{0}=0$ and $\dot{\psi}_{\mu}=0$, so that $\dot{\phi}_{\mu}=-\dot{\phi}_{-\mu}$
and $\delta\tilde{\omega}_{\mu}=-\delta\tilde{\omega}_{-\mu}=D_{3}\mu^{3}/(3\kappa)+\dots$.
Splitting the equations (\ref{simplifiedset}) into real and imaginary
parts we get: 
\begin{eqnarray}
 &  & \alpha_{0}(1+2\sum_{\mu=1}^{L}\alpha_{\mu}\alpha_{-\mu}\sin\psi_{\mu})=f\cos\phi_{0},\nonumber \\
 &  & (\zeta_{0}-2\tilde{N}-2\sum_{\mu=1}^{L}\alpha_{\mu}\alpha_{-\mu}\cos\psi_{\mu}+\alpha_{0}^{2})\alpha_{0}=f\sin\phi_{0},\nonumber \\
 &  & \alpha_{\pm\mu}-\alpha_{\mp\mu}\alpha_{0}^{2}\sin\psi_{\mu}=0,\nonumber \\
 &  & \tilde{\zeta}_{\mu}\alpha_{\mu}-(2\tilde{N}-\alpha_{\mu}^{2})\alpha_{\mu}-\alpha_{-\mu}\alpha_{0}^{2}\cos\psi_{\mu}=0,\label{steadysplit}
\end{eqnarray}
where 
\begin{eqnarray}
 &  & \tilde{\zeta}_{\mu}=(\zeta_{\mu}-2\dot{\phi}_{\mu}/\kappa).
\end{eqnarray}

From the third equation we find that $\alpha_{\mu}=\alpha_{-\mu}$
and $\psi_{\mu}\equiv\psi=\arcsin(1/\alpha_{0}^{2})$. Substituting
this into the first and second equation of (\ref{steadysplit}) we
obtain: 
\begin{eqnarray}
 &  & \tilde{N}=f\alpha_{0}\cos\phi_{0},\nonumber \\
 &  & [\zeta_{0}-2\tilde{N}+\alpha_{0}^{2}+(\tilde{N}-\alpha_{0})\sqrt{1-1/\alpha_{0}^{4}}]\alpha_{0}=f\sin\phi_{0}.\label{eqsetdeg}
\end{eqnarray}
The first equation of (\ref{eqsetdeg}) also follows from the master
equation (\ref{FirstIntegral}). Adding both equations squared we
get the system: 
\begin{eqnarray}
 &  & [\zeta_{0}-\tilde{N}(2+\cos\psi)+\alpha_{0}^{2}(1+\cos\psi)]^{2}\alpha_{0}^{2}+\tilde{N}^{2}/\alpha_{0}^{2}=f^{2},\nonumber \\
 &  & \tilde{\zeta}_{\mu}+\alpha_{\mu}^{2}-2\tilde{N}=\tilde{\zeta}_{nl}=\alpha_{0}^{2}\cos\psi.\label{steadysplit2}
\end{eqnarray}

It follows that $\tilde{\zeta}_{nl}$, which determines the detuning
of the new nonlinear  effective eigenfrequency from the frequency
of oscillations, does not depend on $\mu$. This means that in the
steady state an equidistant comb may be formed if the dispersion of
each mode is compensated by nonlinear pulling with appropriate amplitude
distribution of each excited comb line. This nonlinear dispersion
compensation is possible due to the fact that self phase modulation
is two times smaller than cross phase modulation and is an essential
mechanism for comb formation \cite{Agha2007,Bao2014}.

The last equations of (\ref{steadysplit2}) may be summed over all
excited lines to get: 
\begin{eqnarray}
\tilde{N}=\frac{\zeta_{\Sigma}-\alpha_{0}^{2}+2L\sqrt{\alpha_{0}^{4}-1}}{4L-1},
\end{eqnarray}
where $\zeta_{\Sigma}=\sum_{\mu=-L}^{L}\tilde{\zeta}_{\mu}=(2L+1)\zeta_{0}+d_{2}\sum_{\mu}\mu^{2}$
(where $d_{2}=D_{2}/\kappa$). If excited comb lines are filling all
resonator modes in some interval without omissions the summation may
be done analytically. If substituted in the first equation of (\ref{steadysplit2})
we finally can get an equation for the pump mode magnitude $\alpha_{0}^{2}$
ready for numerical analysis. Substituting $\beta=\alpha_{0}^{2}-\sqrt{\alpha_{0}^{4}-1}$,
$\alpha_{0}^{2}=\frac{\beta^{2}+1}{2\beta}$ then transforms the final
equation to a rational 6-th order equation. Further analysis of these
equations may reveal the properties of the low-noise low-populated
combs observable in experiments.

\subsection{Primary sidebands}

A particular case of the reduced model is a three mode system for
which all the nonlinear terms are in fact considered, i.e. the model
is complete. This system can describe the threshold for primary sideband
hyperparametric oscillations at a $\mu$ times FSR frequency separation
from the pump that can be softly excited when the pump is gradually
tuned into resonance starting from infinite detuning ($\omega_{0}<\omega_{p}$
for $D_{2}>0$).

The three mode system and its stability was analyzed in detail using
numerical analysis of the nonlinear equations \cite{Matsko2005a}
or stability analysis of the linearized equations \cite{Chembo2010}.
Both approaches lead to the same equation following from the reduced
model when $L=1$:

From the nonlinear resonance condition and condition for the first
sideband to emerge ($\alpha_{\mu,\mathrm{th}}^{2}=0,$$\tilde{N}=\alpha_{0}^{2}$)
we obtain: 
\begin{eqnarray}
 &  & (\zeta_{0}-\alpha_{0}^{2})^{2}\alpha_{0}^{2}+\alpha_{0}^{2}=f^{2}\nonumber, \\
 &  & \zeta_{0}+d_{2}{\mu}^{2}=2\alpha_{0}^{2}-\sqrt{\alpha_{0}^{4}-1}.
\end{eqnarray}
It follows the equation determining the possibility to excite the
mode ${\mu}$:

\begin{eqnarray}
 &  & \zeta_{0}=\alpha_{0}^{2}-\sqrt{f^{2}/\alpha_{0}^{2}-1}\nonumber, \\
 &  & \sqrt{f^{2}/\alpha_{0}^{2}-1}-d_{2}{\mu}_{\mbox{th}}^{2}+\alpha_{0}^{2}-\sqrt{\alpha_{0}^{4}-1}=0.
\end{eqnarray}
or using the smallest possible $\alpha_{0}^{2}=1$ when the radical
is real: 
\begin{eqnarray}
{\mu}_{\mathrm{th}}=\left\lfloor \sqrt{(1+\sqrt{f^{2}-1})/d_{2}}\right\rfloor .
\end{eqnarray}
Alternatively, this can be rewritten in physical terms \cite{Herr2012}
as: 
\begin{eqnarray}
{\mu}_{\mathrm{th}}=\left\lfloor \sqrt{\frac{\kappa}{D_{2}}(1+\sqrt{\frac{P_{\mathrm{in}}}{P_{\mathrm{th}}}-1})}\right\rfloor .
\end{eqnarray}
The threshold power $P_{\mathrm{th}}$ corresponds to $|a_{\mathrm{0,th}}|=1$.
The minimum obtainable $\mu_{\mathrm{th}}$ for $P_{\mathrm{in}}=P_{\mathrm{th}}$
is given by: 
\begin{eqnarray}
{\mu}_{\mathrm{th}}=\sqrt{\frac{D_{2}}{\kappa}},
\end{eqnarray}
and depends only on the ratio of cavity decay rate $\kappa$ to second
order cavity dispersion $D_{2}$.

Note that the threshold $\alpha_{\mathrm{0,th}}=1$, determining the
first sideband to appear when tuning the pump into resonance at a
given power (also found in \cite{Chembo2010}), however, does not
correspond to a minimum of input power when hyperparametric generation
may start. This minimum, numerically calculated in \cite{Matsko2005a}
may be found explicitly as $\alpha_{\mathrm{in,min}}=2/\sqrt{3}$.

\section{Dissipative Kerr solitons in optical microresonators}

The formation of stable temporal DKS circulating
in a continuous-wave (CW) driven microresonator with Kerr-nonlinearity
is a remarkable form of optical self-organization. It requires, similarly
to the Kerr-combs described above, a medium with focusing third order
nonlinearity and anomalous group velocity dispersion. Fundamentally
different from the above described Kerr-combs, the formation of temporal
DKS implies a stable pulsed time domain waveform
where the pulse properties are fully determined by the experimental
parameters. The ultra-short soliton pulse duration, typically in the
femto-second regime, implies broadband optical frequency comb spectra
that show the same wide line spacing that is known from the previous
Kerr-combs. Importantly, however, the soliton based comb spectra do
not suffer from the inconsistent subcomb formation and associated
noise processes that are often observed in Kerr-combs. Prior to the
observation in a microresonator, temporal DKS had
already been observed in a several hundred meter long, CW driven fiber
cavity, where additional short laser pulses had been used to stimulate
the soliton formation \cite{Leo2010}. In this context it was speculated
that  DKS could be formed in a microresonator
as well (Note: the possibility of solitons circulating in a dielectric
sphere was already proposed in Ref. \cite{Whitten1997}). Moreover,
Refs. \cite{Matsko2009,Chembo2013} pointed out that the theoretical
formalism so far used by researchers to describe microresonators and
fiber-cavities, i.e. the non-linear coupled mode equations and the
Lugiato-Lefever equation \cite{Lugiato1987} respectively, can be mapped onto each other.
From a historical perspective, it is thus interesting to note that
after the first reports on microresonator Kerr-combs in 2007\cite{DelHaye2007},
it took five years to enter the regime of Kerr solitons
in a microresonator. One reasons for this is the particular thermal
behavior of a microresonator system that had long masked this regime
(cf. Section \ref{sec:Laser Tuning into Solitons}). Moreover, the
approach of stimulating solitons using laser pulses that is successfully
used with long fiber cavities can not easily be transferred to microresonators
due to their much higher finesse (i.e. difficulty of coupling the
pulse to into the resonator). Before presenting experimental observations,
the following section will establish the theoretical formalism that
allows describing dissipative Kerr soliton in a microresonator.

\begin{figure}[h]
\includegraphics[width=1.0\columnwidth]{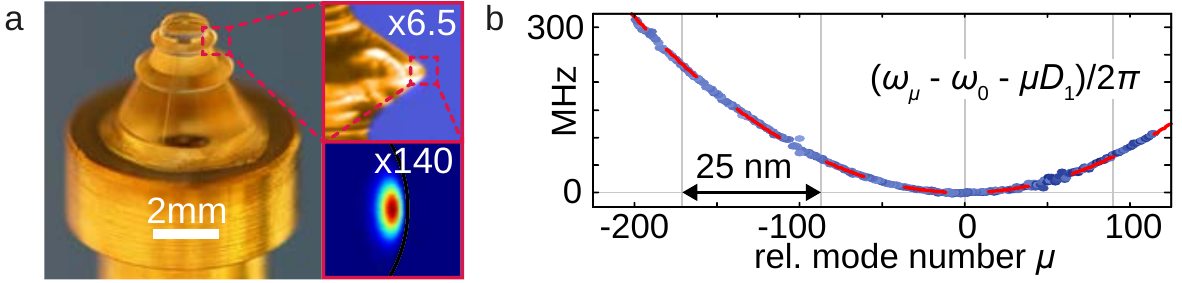}

\protect\protect\protect\protect\caption{\label{fig:Temporal-soliton-formation}Crystalline resonator for dissipative Kerr soliton generation. (a) Crystalline $\mathrm{MgF_{2}}$
resonator and simulated mode profile. (b) Measured anomalous group
velocity dispersion for the resonator shown in panel a. The dispersion
of a resonator can be quantified in terms of the deviation of its
resonance frequencies $\omega_{\mu}$from an equidistant frequency
grid $\omega_{0}+\mu D_{1}$, where $D_{1}/(2\pi)$ is the FSR at
the pump wavelength. An anomalous group velocity dispersion corresponds
to a parabolic curve as shown in panel b.}
\end{figure}

\subsection{Analytical theory of dissipative Kerr solitons\label{sec:AnalyticalTheorySolitons}}

To describe the internal field in a nonlinear microresonator the following
master equation may be used when third and higher order dispersion
terms are neglected \cite{Matsko2011a,Matsko2009}: 
\begin{eqnarray}
\frac{\partial A}{\partial t}-i\frac{1}{2}D_{2}\frac{\partial^{2}A}{\partial\phi^{2}}-ig|A|^{2}A=-\left(\frac{\kappa}{2}+i(\omega_{0}-\omega_{p})\right)A+\sqrt{\frac{\kappa\eta P_{\mathrm{in}}}{\hbar\omega_{0}}}.\label{nls1}
\end{eqnarray}
Here $A(\phi,t)=\sum_{\mu}A_{\mu}e^{i\mu\phi-i(\omega_{\mu}-\omega_{p})t}$
is the slowly varying field amplitude and $\phi\in[0,2\pi]$ is the
angular coordinate inside the resonator with periodic boundary conditions,
$\eta=\kappa_{ex}/\kappa$ is the coupling efficiency with critical
coupling corresponding to $\eta=1/2$. Transforming eq. (\ref{nls1})
to its dimensionless form gives: 
\begin{eqnarray}
i\frac{\partial\Psi}{\partial\tau}+\frac{1}{2}\frac{\partial^{2}\Psi}{\partial\theta^{2}}+|\Psi|^{2}\Psi=(-i+\zeta_{0})\Psi+if.\label{eq:nls}
\end{eqnarray}
Here $\Psi(\tau,\phi)=\sum a_{\mu}(\tau)e^{i\mu\phi}$ is the
waveform, $\theta=\phi\sqrt{\frac{1}{2d_{2}}}$ is the dimensionless longitudinal
coordinate,  and $d_{2}=D_{2}/\kappa$ is the dimensionless dispersion.
Equation (\ref{eq:nls}) is identical to the Lugiato-Lefever equation
\cite{Lugiato1987}, where a transversal coordinate is used instead
of a longitudinal one in our case. It is important to note here, 
that though we used throughout the chapter the term {\it temporal}
to differentiate DCS from better known spacial cavity solitons in nonlinear 
Fabry-Perot etalons, this term may also be confusing. For temporal dissipative
solitons in fibers and fiber loop resonators in equations time is usually 
considered as a ``fast'' variable, describing the form of the pulse, while
the coordinate along propagation is describing slow variations. In the case of
WGM microresonators the opposite mapping is more convenient with ''slow'' time
variation averaged over many roundtrips and fast mapped azimuthal angular 
coordinate. The equation \ref{eq:nls} may also be considered
as a driven, damped Nonlinear Schrödinger Equation (NLS) \cite{Barashenkov1996}.
Analytical solution of the NLS in the form of bright soliton is known
only in case of zero dissipation \cite{Zakharov1972}. An approximate
solution for the dissipative soliton may be found using Lagrangian
perturbation approach \cite{Anderson1983,Mertens2010}. First we introduce
Lagrangian density so that the variation of it over $\Psi^{*}$ (Landau-Euler
equation) leads to an unperturbed NLS equation:

\begin{eqnarray}
\frac{\delta{\cal L}}{\delta\Psi^{*}}\equiv\frac{\partial{\cal L}}{\partial\Psi^{*}}-\frac{\partial}{\partial\tau}\frac{\partial{\cal L}}{\partial\Psi_{\tau}^{*}}-\frac{\partial}{\partial\theta}\frac{\partial{\cal L}}{\partial\Psi_{\theta}^{*}}=0,
\end{eqnarray}

\begin{eqnarray}
{\cal L}=\frac{i}{2}\left(\Psi^{*}\frac{\partial\Psi}{\partial\tau}-\Psi\frac{\partial\Psi^{*}}{\partial\tau}\right)-\frac{1}{2}\left|\frac{\partial\Psi}{\partial\theta}\right|^{2}+\frac{1}{2}|\Psi|^{4}-\zeta_{0}|\Psi|^{2}.
\end{eqnarray}
Now taking into account the perturbed equation (\ref{eq:nls}) we
introduce the dissipative function:

\begin{eqnarray}
\frac{\delta{\cal L}}{\delta\Psi^{*}} & = & {\cal R},\nonumber \\
{\cal R} & = & -i\Psi+if.
\end{eqnarray}
The equations that need to be solved are the following: 
\begin{eqnarray}
\frac{\partial L}{\partial r_{i}}-\frac{d}{d\tau}\frac{\partial L}{\partial\dot{r}_{i}} & = & \int\left({\cal R}\frac{\partial\Psi^{*}}{\partial r_{i}}+{\cal R}^{*}\frac{\partial\Psi}{\partial r_{i}}\right)d\theta,\nonumber \\
L & = & \int{\cal L}d\theta.
\end{eqnarray}
where $r_{i}$ are different possibly time dependent collective coordinates.
Using the ansatz of a stationary ($\frac{\partial\Psi}{\partial\tau}=0$)
soliton $Be^{i\varphi_{0}}{\rm sech}(B\theta)$ (exact for unperturbed
case when $B=\sqrt{2\zeta_{0}}$) with $r_{1}=B$ and $r_{2}=\varphi_{0}$
\cite{Wabnitz1993} we get: 
\begin{eqnarray}
L=-2B\frac{\partial\phi_{0}}{\partial\tau}+\frac{1}{3}B^{3}-2B\zeta_0.
\end{eqnarray}

This yields: 
\begin{eqnarray}
\frac{dB}{d\tau}&=&-2B+\pi f\cos\varphi_{0},\\
\frac{d\phi_{0}}{d\tau}&=&\frac{1}{2}B^{2}-\xi_{0}.
\end{eqnarray}
This leads to the stationary parameters of the soliton attractor \cite{Wabnitz1993}:
\begin{eqnarray}
B&=&\sqrt{2\xi_{0}},\\
\cos\varphi_{0}&=&\frac{2B}{\pi f}=\frac{\sqrt{8\xi_{0}}}{\pi f},\label{eq:B}\\
\zeta_{0}^{\mathrm{max}}&=&\frac{\pi{}^{2}f{}^{2}}{8}.
\end{eqnarray}
An approximate single soliton solution which accounts for a flat CW
background is then given by 
\begin{eqnarray}
 & \Psi=\Psi_{0}+\Psi_{1}\simeq\Psi_{0}+Be^{i\varphi_{0}}{\rm sech}(B\theta),\label{eq:singlesoliton}
\end{eqnarray}
The constant CW background $\Psi_{0}$ may be found by inserting $\Psi_{0}$
into eq. (\ref{eq:nls}) as the lowest branch \cite{Barashenkov1996}
of the solution of 
\begin{eqnarray}
 & (|\Psi_{0}|^{2}-\zeta_{0}+i)\Psi_{0}=if,
\end{eqnarray}
which, eventually, results for $\zeta_{0}>\sqrt{3}$ (bistability
criterion) and large enough detunings $f^{2}<\frac{2}{27}\zeta_{0}(\zeta_{0}^{2}+9)$
in: 
\begin{eqnarray}
|\Psi_{0}|^{2} & =&\frac{2}{3}\zeta_{0}-\frac{2}{3}\sqrt{\zeta_{0}^{2}-3}\cosh\!\left(\frac{1}{3}\,\,\mathrm{arcosh}\!\left(\frac{2\zeta_{0}^{2}+18\zeta_{0}-27f^{2}}{2(\zeta_{0}^{2}-3)^{2/3}}\right)\right),\nonumber \\
\Psi_{0} & =&\frac{if}{|\Psi_{0}|^{2}-\zeta_{0}+i}\simeq\frac{f}{\zeta_{0}^{2}}-i\frac{f}{\zeta_{0}}.\label{eq:Psi0}
\end{eqnarray}
Extending (\ref{eq:singlesoliton}) to the case of multiple solitons
inside the resonator gives 
\begin{eqnarray}
 & \Psi(\phi)\simeq\Psi_{0}+\left(\frac{4\zeta_{0}}{\pi f}+i\sqrt{2\zeta_{0}-\frac{16\zeta_{0}^{2}}{\pi^{2}f^{2}}}\right)\sum_{j=1}^{K}\,\mathrm{sech}(\sqrt{\frac{\zeta_{0}}{d_{2}}}(\phi-\phi_{j})).\label{waveform}
\end{eqnarray}
This ansatz automatically translates to discrete possibles states
(steps) in nonlinear resonant response. These steps are an important
signature of soliton Kerr combs and are observed in the experiments
(see below). The height of steps in the intracavity power can be found
by averaging the waveform amplitude (eq. \ref{waveform}) squared
over one roundtrip for different numbers $K$ of solitons: 
\begin{eqnarray}
\overline{|\Psi|^{2}} & =|\Psi_{0}|^{2}+K\frac{1}{2\pi}\int_{0}^{2\pi}(\Psi_{1}^{2}+\Psi_{0}\Psi_{1}^{*}+\Psi_{1}\Psi_{0}^{*})d\phi\nonumber \\
 & \simeq\frac{f^{2}}{\zeta_{0}^{2}}+K\frac{2}{\pi}\sqrt{d_{2}\zeta_{0}}.
\end{eqnarray}
The Fourier transform ($\mathcal{F}$) of a hyperbolic secant soliton is again a hyperbolic
secant: 
\begin{eqnarray}
\Psi(\mu)=\mathrm{\mathcal{F}}\left[\sqrt{2\zeta_{0}}\,\mathrm{sech}\left(\sqrt{\frac{\zeta_{0}}{d_{2}}}\phi\right)\right]=\sqrt{d_{2}/2}\,\mathrm{sech}\left(\frac{\pi\mu}{2}\sqrt{\frac{d_{2}}{\zeta_{0}}}\right).
\end{eqnarray}
Using the relation for the optical frequency $\omega=\omega_{p}+\mu D_{1}$
and the time $t=\phi/D_{1}$ the spectral envelopes and the soliton
waveform can be rewritten: 
\begin{eqnarray}
\Psi(\omega-\omega_{p})=\sqrt{d_{2}/2}\,\mathrm{sech}((\omega-\omega_{p})/\Delta\omega)\ \ \mathrm{with}\ \Delta\omega=\frac{2D_{1}}{\pi}\sqrt{\frac{\zeta_{0}}{d_{2}}},
\end{eqnarray}
and 
\begin{eqnarray}
\Psi(t)=\sqrt{2\zeta_{0}}\,\mathrm{sech}(t/\Delta t),\ \ \mathrm{with}\ \Delta t=\frac{1}{D_{1}}\sqrt{\frac{d_{2}}{\zeta_{0}}}.
\end{eqnarray}
The minimal possible soliton duration can be found by using $\zeta_{0}^{\mathrm{max}}$
(eq. \ref{eq:B}) in the above equation for $\Delta t$: 
\begin{eqnarray}
\Delta t_{\mathrm{min}}=\frac{1}{\pi D_{1}}\sqrt{\frac{\kappa D_{2}n_{0}^{2}V_{\mathrm{eff}}}{\eta P_{\mathrm{in}}\omega_{0}cn_{2}}}.
\end{eqnarray}
This equation can be recast in form of the group velocity dispersion
$\beta_{2}=\frac{-n_{0}}{c}D_{2}/D_{1}^{2}$, the nonlinear parameter
$\gamma=\frac{\omega}{c}\frac{n_{2}}{{\cal A_{\mathrm{eff}}}}$ (for
simplicity we assume critical coupling $\eta=1/2$ and on resonant
pumping): 
\begin{eqnarray}
\Delta t_{\mathrm{min}}=\frac{2}{\sqrt{\pi}}\sqrt{\frac{-\beta_{2}}{\gamma{\cal {F}P_{\mathrm{in}}}}},
\end{eqnarray}
where denotes the finesse ${\cal F}=\frac{D_{1}}{\kappa}$ of the
cavity. Note that the values $\Delta\omega$ and $\Delta t$ need
to be multiplied by a factor of $2\,\mathrm{arccosh}(\sqrt{2})=1.763$
to yield the FWHM of the sech$^{2}$-shaped power spectrum and pulse
intensity, respectively.

For the time bandwidth product (TBP) we find $\Delta t\cdot\Delta\omega=2/\pi$
or, when considering the FWHM of spectral and temporal power (in units
of Hz and s), $\mathrm{TBP}=0.315$.

In the case of $K$ multiple solitons inside the cavity there is a
more structured optical spectrum $\Psi_{k}$, resulting from interference
of single soliton spectra $\Psi_{j}(\mu)$, where the relative phases
of these spectra are determined by the positions $\phi_{j}$ of individual
solitons: $\Psi_{K}(\mu)=\Psi(\mu)\sum_{j=1}^{K}\mathrm{e^{i\mu\phi_{j}}}$.
The line-to-line variations can be high, however, the overall averaged
spectrum still follows the single soliton shape and is proportional
to $\sqrt{K}\Psi(\mu)$.

\section{Signatures of  dissipative Kerr soliton formation in crystalline
resonators\label{sec:Signatures of Temporal Solitons}}

\begin{figure}[h]
\includegraphics[width=1.0\columnwidth]{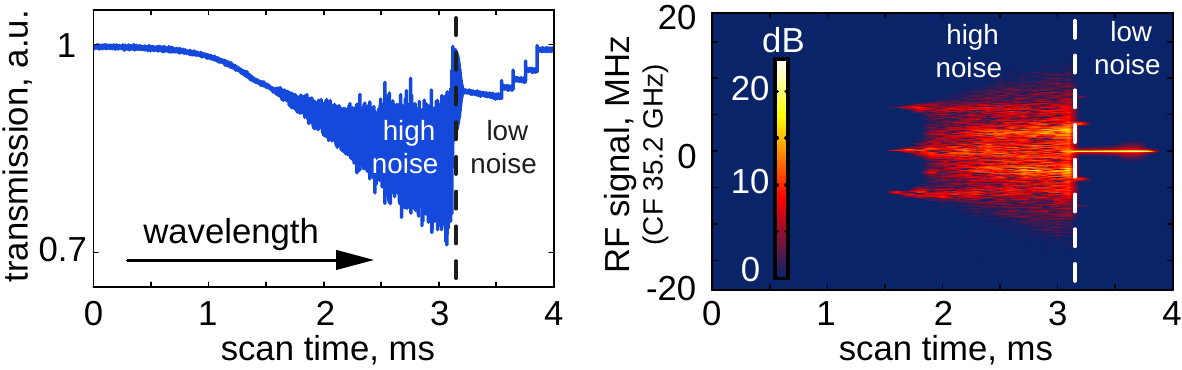}
\protect\protect\protect\protect\caption{\label{fig:Observation-of-discrete-Steps-1}Signatures of soliton
formation. (a) A staircase-like step structure in the pump laser transmission
indicates the formation of several temporal DKS in
the microresonator. High intensity noise is apparent in the transmission
signal prior to soliton formation. (b) The formation of the solitons
is associated with a transition to low noise. This is evidenced here
by the transition of a broad RF beatnote (cf. Figure \ref{fig:Emergence-of-Noise})
to a single narrow-band signal. (CF: center frequency)}
\end{figure}

The generation of DKS in microresonators
is associated with several characteristic signatures that can already
be observed during a transient generation of the solitons. Here transient
generation refers to the situation where the pump laser is scanned
(with increasing wavelength) through the resonance of the microresonator.
The described signatures are particularly useful from a practical
perspective as they allow for rapid sample characterization. Depending
on the detuning during the laser scan the system can enter different
nonlinear regimes including, parametric sideband generation with Turing
patterns\cite{Coillet2013}, Kerr-comb formation, breather solitons
\cite{Matsko2012} and stable dissipative Kerr soliton generation.
One signature, though not unambiguously connected to soliton formation,
is a transition from high intensity noise and and high phase noise
(related to Kerr-subcomb formation) to a low noise state. While scanning
the laser, the noise can be monitored by e.g. recording the intensity
noise in the generated spectrum or by detecting the beatnote signal
between the generated optical lines as shown in Figure \ref{fig:Observation-of-discrete-Steps-1}b.

\begin{figure}[h]
\includegraphics[width=1.0\columnwidth]{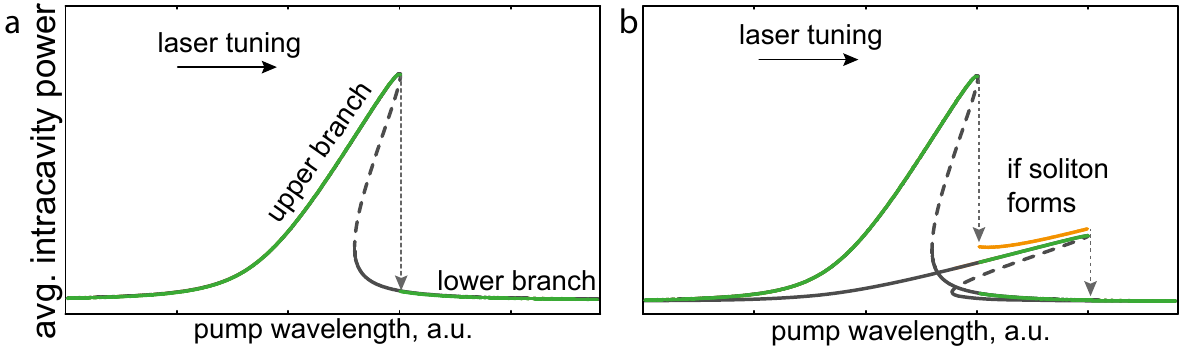}
\protect\protect\protect\protect\caption{\label{fig:Stability-of-soliton-Curve}Stability of soliton states:
Considering only the Kerr-nonlinear resonance shift the intracavity
power can be described by bistability curves where the upper branch
solution corresponds to high and the lower branch solution to low
intracavity power. When tuning into the resonance with decreasing
optical frequency (increasing wavelength) the intracavity power follows
the upper branch of the Kerr-bistability curve. After the transition
to a soliton state the major fraction of the pump light is described
by the lower branch of the bistability curve. The fraction of the
pump light that propagates with the soliton inside the microresonator
experiences a larger phase shift and is effectively blue detuned on
the upper branch of another bistability curve. The extend of the ’soliton
bistability curve’ towards longer wavelength depends on the peak power
of the solitons (i.e. the maximal nonlinear phase shift), the relative
height of the curve depends on the relative fraction of the pump light
that is affected by the high peak power soliton. The overall intracavity
power can be inferred by adding the bistability curves resulting in
the black curve.}
\end{figure}

\begin{figure}[h]
\includegraphics[width=1\columnwidth]{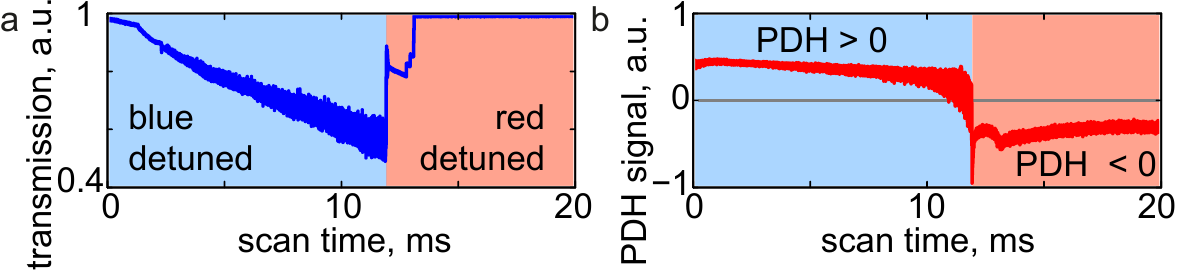}
\protect\protect\protect\protect\caption{\label{fig:Laser-detuning-of-Soliton-State}Laser detuning and soliton
formation. Panel (a) shows the transmitted power and a series of steps
associated with multiple temporal Kerr solitons. The background
shading indicates the laser detuning that is derived from a Pound-Drever-Hall
(PDH) error signal shown in panel (b). The soliton formation coincides
with the transition to red detuning (indicated by a sign change in
the PDH error signal). Note that the additional Kerr-frequency shift
due to the high peak power soliton does not significantly impact the
PDH error signal as its relative contribution is small. }
\end{figure}

A more unique and easily observable signature of soliton formation
in optical microresonators is the occurrence of discontinuous staircase-like
steps in the transmitted pump laser power while the laser is continuously
scanned across the optical resonance (cf. Figure \ref{fig:Observation-of-discrete-Steps-1}a
) \cite{Herr2013}%
\footnote{Discontinuous steps can also be observed (typically with a higher
signal-to-noise ratio) in the power of the nonlinearly generated light.%
}. When operating a microresonator a lower pump laser transmission
implies a higher average intracavity power. The emergence of the steps
may be explained by referring to the cavity bistability curve that
describes the power inside the cavity in dependence of the pump laser
detuning. When approaching the cavity resonance from the blue detuned
side (laser frequency is higher than the resonance frequency and correspondingly
laser wavelength is shorter than the resonance wavelength) the intracavity
power will increase. This increase of intracavity power will then,
due to the Kerr-nonlinearity, effectively shift the resonance frequency
towards longer wavelength. As the laser detuning is further decreased
the intracavity power will steadily increase (upper branch on the
bistability curve), until the highest possible power is reached at
the point of effective zero-detuning. Beyond this point (now effectively
red-detuned) the intracavity-power will steeply drop to a much lower
value (lower branch on the bistability curve) as the Kerr-nonlinear
resonance shift vanishes (cf. Figure \ref{fig:Stability-of-soliton-Curve}a).
Exactly at this transition to red detuning temporal DKS
can emerge from the noisy intracavity waveform (cf. Figure \ref{fig:Laser-detuning-of-Soliton-State}).
In this case the formation of a high peak power soliton pulse will
cause an additional Kerr-frequency shift such that the part of the
pump light that sees the soliton pulse (i.e. copropagating with the
soliton inside the cavity) follows now a second bistability curve.
\ref{fig:Stability-of-soliton-Curve}b. The result is a second power
drop at a larger detuning when the soliton vanishes.

The combined signal of power drops occurring at different laser detunings
leads to the overall staircase-like step structure in the transmitted
power. If a higher number of soliton pulses is generated, a series
of multiple steps can be observed (during a rapid scan not all solitons
disappear at the exact same detuning). The exact number of generated
solitons differs from one laser scan to the other as the solitons
emerge from a noisy state. It is interesting to note that the transition
to effective red-detuning has not been considered in previous work
as it is usually associated with only low intracavity power \cite{Carmon2004}.
Section \ref{sec:Laser Tuning into Solitons} will describe how stable
operation in the soliton regime can be achieved despite the steps
in the intracavity power that can affect the resonator temperature.

\section{Laser tuning into the Kerr soliton states\label{sec:Laser Tuning into Solitons}}

For stable operation of a microresonator system thermal self-locking
\cite{Carmon2008,Li2014} is a widely used method. Here the resonator
self-locks itself thermally to the pump laser frequency. Besides being
a simple and convenient method, this allows for a fixed pump laser
frequency during steady state operation (instead of a pump laser that
follows drifts in the resonator). This is advantageous for frequency
comb operation, where the pump laser constitutes one of the comb lines
that should remain fixed in frequency. The mechanism of thermal self-locking
is somewhat analogous to the Kerr-nonlinear bistability curve and
gives rise to a so-called thermal triangle, that is known from the
earliest reports of whispering-gallery mode microresonators\cite{Braginsky1989,Ilchenko1992a}.
In a resonator with positive thermo-refractive coefficient (as in
$\mathrm{MgF_{2}}$ or $\mathrm{Si_{3}N_{4}}$) thermal locking is
achieved by blue detuned pumping. In this regime an increase of pump
wavelength (or equivalently a decrease of resonance wavelength) implies
a higher intracavity power and higher rate of absorptive heating.
As a consequence of this heating and mostly due to the thermo-optic
effect the resonance wavelength shifts towards longer wavelength thus
stabilizing the relative detuning between pump laser and resonator.
Exactly this behavior is present during a soliton state. This can
be seen e.g. in Figure \ref{fig:Observation-of-discrete-Steps-1}
where (within a particular soliton state) an increase in pump wavelength
causes an increase in intracavity power (decrease of transmission).

\begin{figure}[h]
\includegraphics[width=1.0\columnwidth]{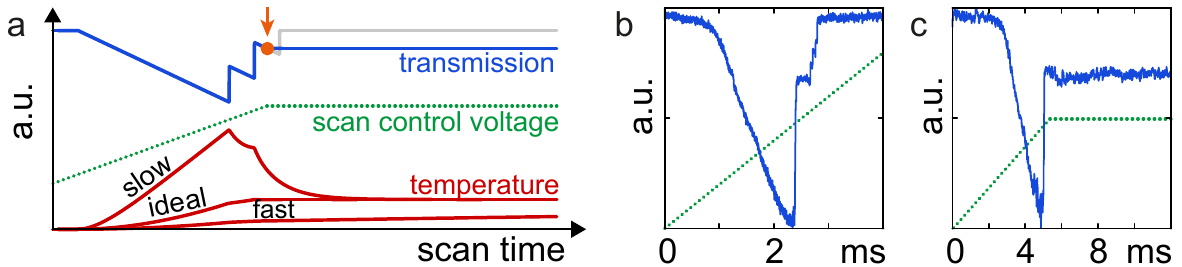}

\protect\protect\protect\protect\caption{\label{fig:Laser-tuning-technique}Generation of stable solitons via
laser tuning. (a) Illustration of laser transmission, scan control
voltage (corresponding to pump wavelength) and resonator temperature
for slow, ideal and fast laser tuning speed. In the ideal case the
resonator temperature does not change once the desired soliton state
is reached and the laser scan can be stopped for stable operation.
(b) Regular scan of the pump laser over a resonance showing a 'soliton
step'. (c) The laser tuning method allows to stop the laser scan once
the soliton is generated. Once generated in this manner, the soliton
circulates stably inside the microresonator.}
\end{figure}

In order to stably reach the regime of thermal locking of a soliton
state it is important that the transition to the soliton state does
not induce a change of the resonator temperature (despite the fact
that the intracavity power changes), which would directly destabilize
the system. One solution to this problem is to tune the pump laser
from a far-blue detuned position into the soliton state with a tuning
speed chosen such, that the ``integrated absorptive heating''
over the tuning time corresponds to the heating rate in the soliton
state. In this case the temperature does not change significantly
when the soliton state is reached. Thermal locking then ensures the
stable operation of the soliton state (for many hours), without actuation
on the pump laser frequency, as illustrated in Figure \ref{fig:Laser-tuning-technique}.
It is important to note that the solitons emerge from intrinsic noise
present in the cavity prior to reaching the soliton state. As a consequence
the initial number of solitons generated via the laser tuning method
varies randomly from shot to shot. However, the experimental conditions
can often be adjusted such that a certain number of solitons (e.g.
one soliton) is generated with high probability. The maximal number
of soliton that can be generated via the laser tuning method is related
to the dispersion of the cavity and grows with the applied pump power
\cite{Herr2013}. For typical parameters only a few, maximally up
to approximately 10 or 20, solitons are generated. As the exact number
depends on a random process the electronically controlled laser tuning
method is usually repeated several times until it yields the desired
number of solitons. Note that the laser tuning method does not require
an

additional external stimulation (e.g. a laser pulse), as it is the
case in fiber cavities. The following section will characterize the
solitons generated via the described laser tuning method.

\section{Simulating soliton formation  in\\ microresonators \label{sec:Simulating Soliton Formation}}

Dissipative temporal Kerr solitons in a microresonator can be simulated
by integrating the coupled mode equations (\ref{coupled_modes_eq})
in time similar to work related to Kerr-combs by \cite{Chembo2010,Chembo2010b,Chembo2010d}.
We use an adaptive Runge-Kutta method of the 5-th order, which together
with a transformed system of equations without trigonometric functions
and explicit time dependence allows achieving fast simulation. The
simulation of hundreds of optical lines  allows for accurate modeling
of realistic experimental systems. This includes in particular the
beatnote between the optical lines (i.e. the repetition rate), the
envelope of the optical spectrum as well as instant amplitudes, phases
and frequencies of all modes. The sum of the nonlinear mixing terms
is efficiently calculated in the Fourier-domain\cite{Hansson2014}.
An additional self-steepening term is added when simulating broad
combs in $\mathrm{Si_{3}N_{4}}$ rings to better model the behavior
of few-cycles pulses\cite{Lamont2013}. In simulations, similar to
the experiment, solitons can be generated via tuning the pump laser
into resonance. The number of generated solitons depends on the noisy
intracavity waveform preceding the soliton state. To allow for deterministic
simulations of a certain number of solitons, the respective number
of sech-pulses can be seeded as an initial waveform while the simulation
is started directly with a detuning that allows for stable solitons.
It is worthwhile noting that the underlying coupled mode equations
readily allow for the simulation of complex and irregular mode structures
as often encountered in real microresonators (cf. Section \ref{sec:Mode structure and soliton formation}).
Based on our simulations we developed an open-source graphical user
interface using \texttt{MATLAB}  \cite{Lihachev2014} that allows numerical
simulation of Kerr-comb and soliton generation in microresonators.

\begin{figure}[]
\includegraphics[width=1.0\columnwidth]{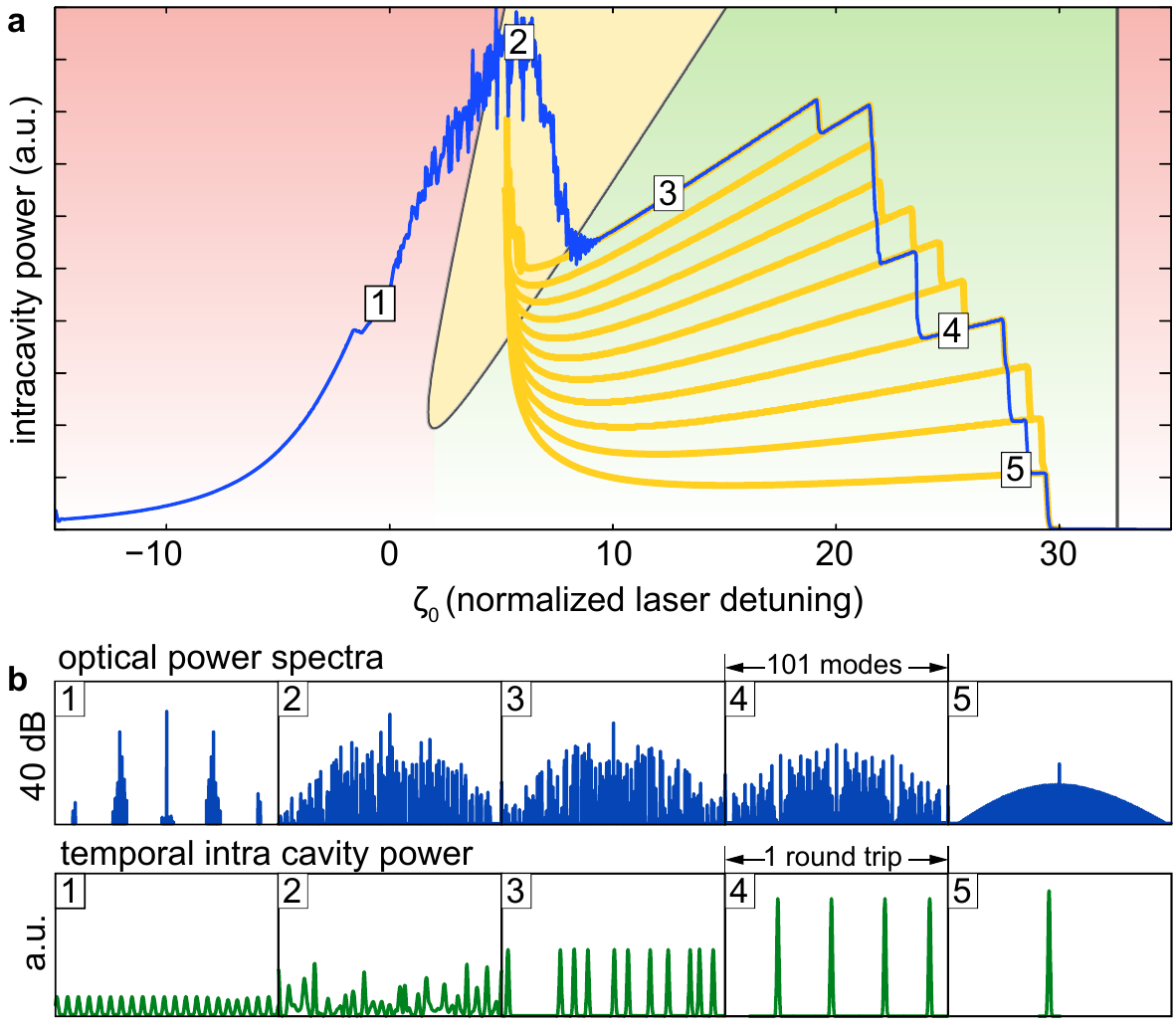}

\protect\protect\protect\protect\caption{Numerical simulations of dissipative Kerr soliton formation in a crystalline microresonator. (a) Intracavity power (corresponding
to the transmission signal in Figure \ref{fig:Observation-of-discrete-Steps-1}a
when mirrored horizontally) during a simulated laser scan (101 simulated
modes) over a resonance in a microresonator. The step features are
clearly visible. The light gray lines trace out all possible states
of the system during the scan. The unshaded area corresponds to the
area where temporal DKS can exist, the light shaded
area allows for breather solitons with a time variable, oscillating
envelope; no solitons can exist in the dark shaded area. b. Optical
spectrum and intracavity intensity for different detuning values (1-5)
in the laser scan. }
\end{figure}

\section{Characterization of DKS in crystalline
microresonators\label{sec:Characterizing temporal solitons}}

\begin{figure}[H]
\includegraphics[width=1\columnwidth]{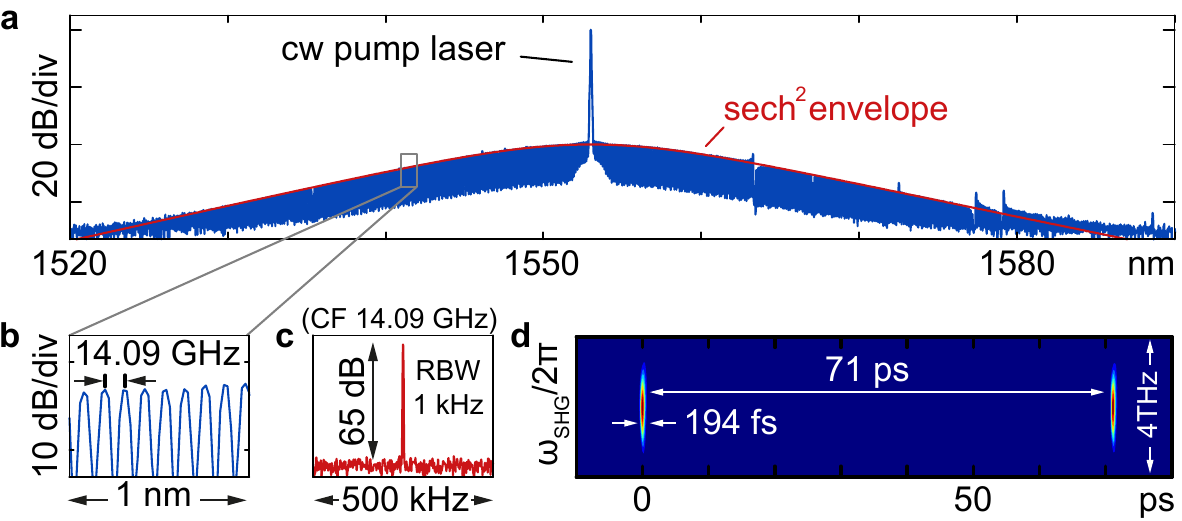}

\protect\protect\protect\protect\caption{\label{fig:Single-Temporal-Soliton-Frog}Characteristics of a temporal
dissipative Kerr soliton in a crystalline resonator. (a) Optical spectrum
showing the characteristic $\mathrm{sech}^2$-envelope. (b) Magnified part
of the spectrum, resolving the individual comb lines of which it is
composed. (c) The low noise radio-frequency beatnote at 14.09 GHz
corresponds to the comb line spacing and the soliton pulse repetition
rate. (d) SHG-FROG trace revealing femto-second pulse duration. The
pulse to pulse separation of 71 ps corresponds to the pulse repetition
rate of 14.09 GHz and the pulse duration of 194 fs can be inferred
(in agreement with the spectral width). (RBW: resolution bandwidth,
CF: center frequency)}
\end{figure}

The ability to stably tune into soliton states as described in the
Section \ref{sec:Laser Tuning into Solitons} allows for characterization
of the solitons' spectral and temporal properties in crystalline microresonators
$\mathrm{MgF_{2}}$. Precisely as predicted by theory, the single
soliton state has a \emph{sech}-squared spectral envelope from which a pulse
duration in the femto-second regime can be inferred. The latter can
be confirmed via direct temporal characterization. Here, frequency
resolved optical gating using a nonlinear Michelson-interferometer
(SHG-FROG)\cite{Kane1993} is used, cf. Figure \ref{fig:Single-Temporal-Soliton-Frog}.
This method also allows for the reconstruction of the complex pulse
envelope\cite{Kane2008}. The high pulse repetition rate inherent
to microresonators allows the FROG interferometer delay to extend
over several pulse round-trip times, i.e. consecutive pulses can be
cross-correlated.

\begin{figure}[h]
\includegraphics[width=1.0\textwidth]{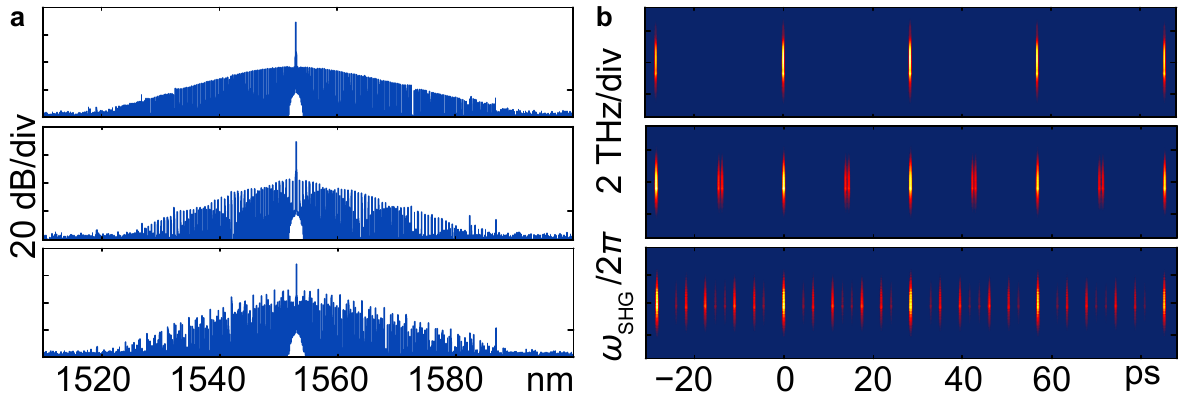}

\protect\protect\protect\protect\caption{\label{fig:FROG-Multiple-Soliton}Spectral and temporal characterization
of  multi-soliton states in a crystalline resonator. (a) Optical spectra
of a single soliton (top) and two multi-soliton states (middle, bottom).
(b) SHG-FROG traces corresponding to (a).}
\end{figure}

Besides stable single soliton states, also stable multi-soliton states can
be generate using the laser tuning method. Figure \ref{fig:FROG-Multiple-Soliton}
compares the optical spectra and SHG-FROG traces of a single and two
different multi-soliton states. A salient feature of multi-soliton
states is the modulated spectral envelope arising from the interference
of the individual soliton. Based on the spectral envelope the relative
positions of the solitons can be reconstructed (cf. Section \ref{sec:SiNSolitons}).
Once a multi-soliton state is generated inside the cavity, the envelope
pattern does not change over several hours, implying that the temporal
separation between the solitons in the microresonator does not change
on this time scale.

\section{Resonator mode structure and soliton\\ formation\label{sec:Mode structure and soliton formation} }

\begin{figure}[H]
\includegraphics[width=1.0\textwidth]{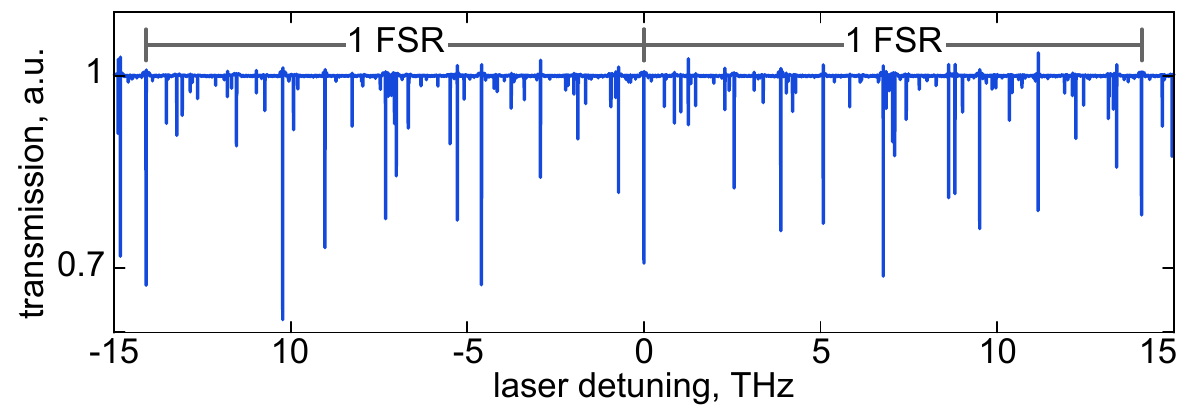}

\protect\protect\protect\protect\caption{\label{fig:Transmission-spectrum-of-Microresonator}Transmission spectrum
of a $\mathrm{MgF_{2}}$ crystalline microresonator with a FSR of
approximately $14.09$ $\mathcal{\mathrm{GHz}}$. The upward transmission
spikes (values $>1$) result from cavity-ringdown. Frequency comb
assisted diode laser spectroscopy ensures precise calibration of the
laser detuning (MHz level).}
\end{figure}

\begin{figure}[H]
\includegraphics[width=1.0\columnwidth]{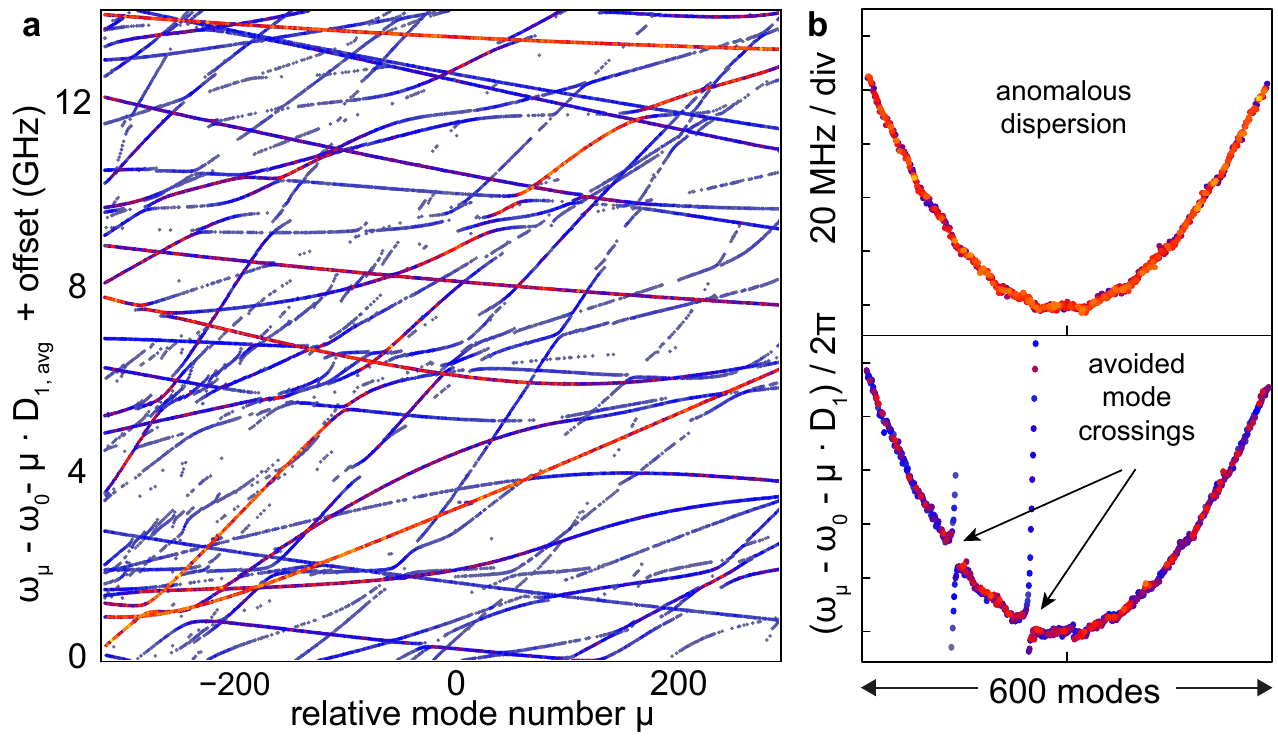}

\protect\protect\protect\protect\caption{\label{fig:Echelle-Plot}Mode structure of a $\mathrm{MgF_{2}}$resonator
with a FSR of 14.09 GHz. (a) Two-dimensional Echelle-type representation
where for all measured mode families the deviation $\omega_{\mu}-\omega_{0}-\mu\cdot D_{1,\mathrm{avg}}$
of the resonance frequency $\omega_{\mu}$ from an equidistant $D_{1,\mathrm{avg}}$
-spaced frequency grid (D1,avg is an approximate average FSR of all
modes) is shown (plus some offset) in function of the mode number
$\mu$. Dots forming a continuous line represent a particular mode
family. Different free spectral ranges correspond to different slopes
of the lines, whereas dispersion and variation of the FSR show as
curvature and bending of the lines. The dispersion can be strongly
affected by mode crossings. (b) Two mode families have been extracted
from the data set shown in (a). The upper one is characterized by
an anomalous dispersion, the lower one exhibits two avoided mode crossings
that induce deviations from the anomalous dispersion. }
\end{figure}

\begin{figure}[H]
\includegraphics[width=1.0\textwidth]{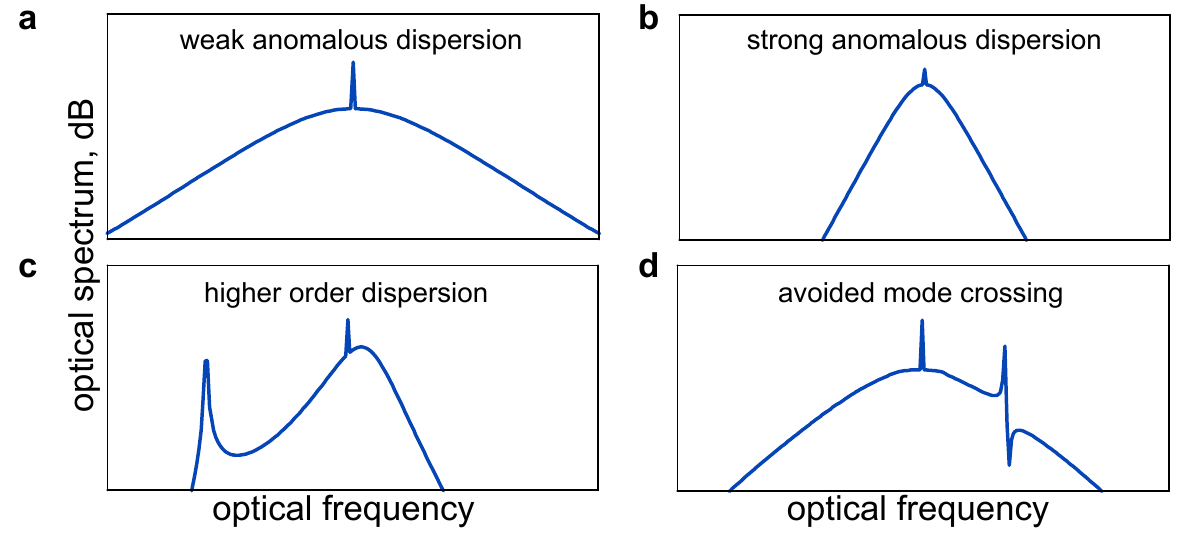}

\protect\protect\protect\protect\caption{\label{fig:ModeCrossingSpectrum}Mode-structure and spectral envelope.
(a) Typical sech-squared envelope for the case of weak (a) and strong
anomalous dispersion (b). (c) Higher order dispersion (such as non
zero $D_{3}$ ) leads to an asymmetric spectrum, dispersive wave emission
(cf. Section \ref{sec:SiNSolitons}) and a shift of the spectral soliton
peak intensity away from the pump laser (soliton recoil). (d) Avoided
mode crossings manifest themselves in a characteristic spectrally
local variation of the spectrum.}
\end{figure}

The generation of bright temporal DKS requires a
focusing nonlinearity and an anomalous group velocity dispersion (GVD),
i.e. a linear growth of the resonator's FSR with the mode number.
In this case the growth of the FSR can be compensated by nonlinear
mode pulling, i.e. nonlinear optical shifts of the resonance frequency.
 In principle most microresonators fulfill this requirement in the
(near-) infrared spectral regime. As opposed to optical fiber systems,
microresonators, however, can exhibit a rich and complex optical mode
structure. Some modes are affected by higher order dispersion or are
even characterized by normal GVD. Moreover, fabrication defects and
asymmetries can lead to modal coupling and distortions of the mode
spectrum due to avoided mode crossings. While temporal dissipative Kerr
solitons are a single-mode phenomenon, the structure of the respective
mode-family can be altered by the presence of the other modes. The
understanding of the impact and influence of the mode-structure is
a decisive step in understanding DKS in optical microresonators. 
In Ref. \cite{Herr2014} this is studied
experimentally and with the help of the numerical tools described
in Section \ref{sec:Simulating Soliton Formation}. First, the mode
structure of the microresonator is precisely measured using frequency
comb assisted diode laser spectroscopy \cite{DelHaye2009a}. Figure
\ref{fig:Transmission-spectrum-of-Microresonator} shows a fraction
of the recorded transmission spectrum of a crystalline resonator,
revealing several mode families. The full data set is represented
in Figure \ref{fig:Echelle-Plot}a. Individual mode families and their
dispersion can be extracted from the full data set. An example of
a mode family with almost perfect anomalous GVD is shown in Figure
\ref{fig:Echelle-Plot}b, top. In contrast, Figure \ref{fig:Echelle-Plot}b,
bottom shows a another mode family where the globally anomalous GVD
is locally modified by two avoided mode crossing. While in the first
case solitons can be generated, this is not possible in the latter.
The hypothesis that avoided mode crossing (if they occur in the spectral
vicinity of the pumped mode) inhibit soliton formation is confirmed
by numerical simulation \cite{Herr2014}. Moreover, strong higher
order dispersion can prevent soliton formation. Importantly, moderate
contribution of higher order dispersion and avoided mode crossings
further away from the pump wavelength do not inhibit soliton formation
but manifest themselves in the optical spectrum as qualitatively illustrated
in Figure \ref{fig:ModeCrossingSpectrum}. Indeed, similar structures
can be observed in the experimentally generated soliton spectra e.g.
in Figure \ref{fig:Single-Temporal-Soliton-Frog}.

\section{Using dissipative Kerr solitons to count the cycles of light}

\begin{figure}[H]
\includegraphics[width=1.0\columnwidth]{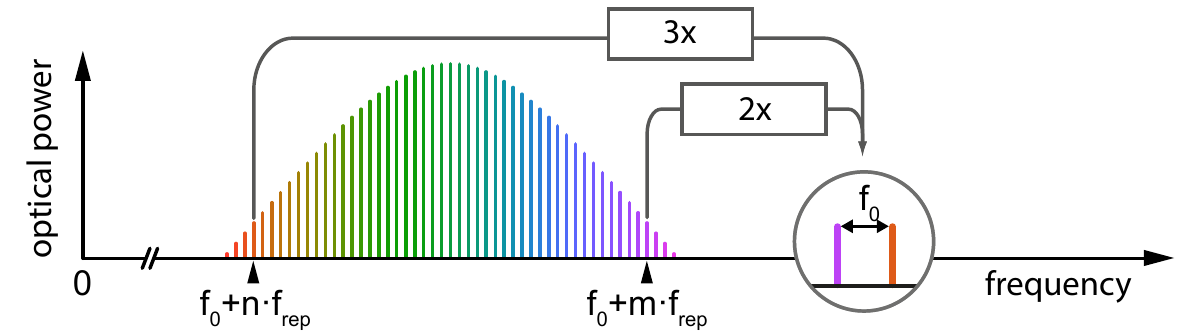}

\protect\protect\protect\protect\caption{\label{fig:Experimental-scheme-2f-3f}Illustration of $2f$-$3f$
self-referencing. If the frequency comb spectrum spans more than two
thirds of an octave, the second and third harmonics of blue and red
wings of the spectrum can be overlapped. The difference frequency
beatnote between the two harmonics yields the comb's offset frequency
$f_{0}$. }
\end{figure}

\begin{figure}[H]
\includegraphics[width=1\columnwidth]{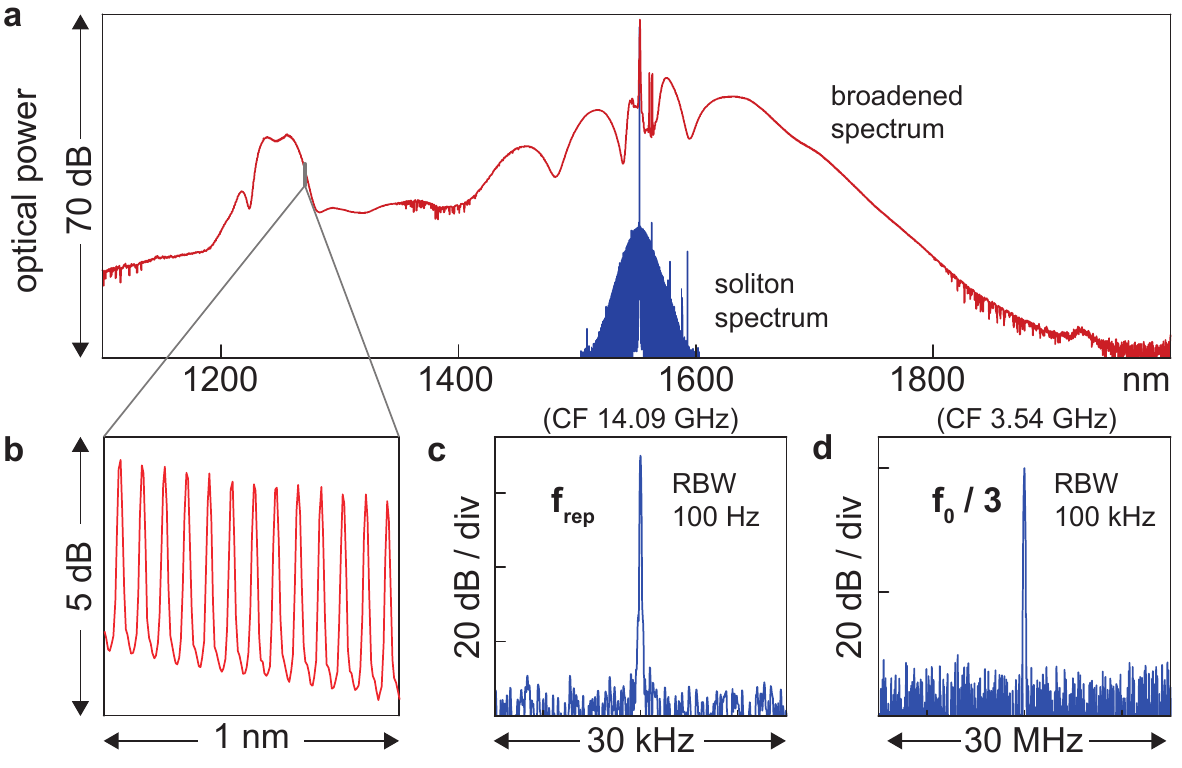}

\protect\protect\protect\protect\caption{\label{fig:fceo beatnote}Counting the cycles of light by self-referencing
a soliton based frequency comb. (a) Soliton spectrum and nonlinearly
broadened spectrum. The broadened spectrum spans more than two thirds
of an octave and allows for $2f-3f$ self-referencing. (b) Magnified
part of the broadened spectrum. The line spacing is the same as for
the soliton spectrum (14.09 GHz). (c) Pulse repetition rate beatnote
measured via direct photo-detection of the comb spectrum. (d) Offset
frequency signal measured via a modified $2f-3f$ self-referencing
technique, where two transfer laser are used for signal enhancement
(see Ref. \cite{Jost2014} for details). (RBW: resolution bandwidth,
CF: center frequency)}
\end{figure}

The observation of Kerr solitons in microresonators, and the associated
generation of ultra-short femto-second optical pulses, provides an
opportunity to count the cycles of light using an optical microresonator,
i.e. to link phase coherently an optical laser frequency to the radio
frequency domain. In order to achieve such a radio to optical frequency
link\cite{Diddams2000} self-referencing has to be employed; a process
where the comb's offset frequency $f_{0}$ (cf. Figure \ref{fig:CombsGeneral})
is measured. To this end, an optical spectrum of sufficient spectral
span is required that spans an either a full octave when using the
$1f$-$2f$ self-referencing technique or 2/3 of an octave for the
$2f$-$3f$ self-referencing method\cite{Telle1999,Reichert1999,Udem2002}
(cf. Figure \ref{fig:Experimental-scheme-2f-3f}). While the soliton
spectrum generated in a crystalline resonator is not sufficiently
broad by itself, the short pulse duration enables (after power amplification
in an erbium-doped fiber amplifier) spectral broadening \cite{Dudley2006}
in highly-nonlinear optical fiber. Figure \ref{fig:fceo beatnote}a
shows a soliton spectrum that has been broadened in this way \cite{Jost2014}.
Here, $2f$-$3f$ self-referencing is implemented allowing for detection
of the soliton comb's offset frequency $f_{0}$ (cf. \ref{fig:fceo beatnote}d).
Together, with the comb line spacing (i.e. the pulse repetition rate)
$f_{\mathrm{rep}}$ that can readily measured via direct photo-detection
(cf. \ref{fig:fceo beatnote}c ), the frequencies constituting the
soliton comb are fully defined. In this way, the temporal dissipative
Kerr soliton allows linking optical frequencies (e.g. the pump laser wavelength
or any other comb line) to the radio (or microwave) frequencies $f_{0}$
and $f_{\mathrm{rep}}$.

\clearpage{}

\section{Dissipative Kerr solitons and soliton induced\\ Cherenkov radiation in a $\mathrm{Si_{3}N_{4}}$
photonic chip\label{sec:SiNSolitons}}

\begin{figure}[H]
\includegraphics[width=1.0\columnwidth]{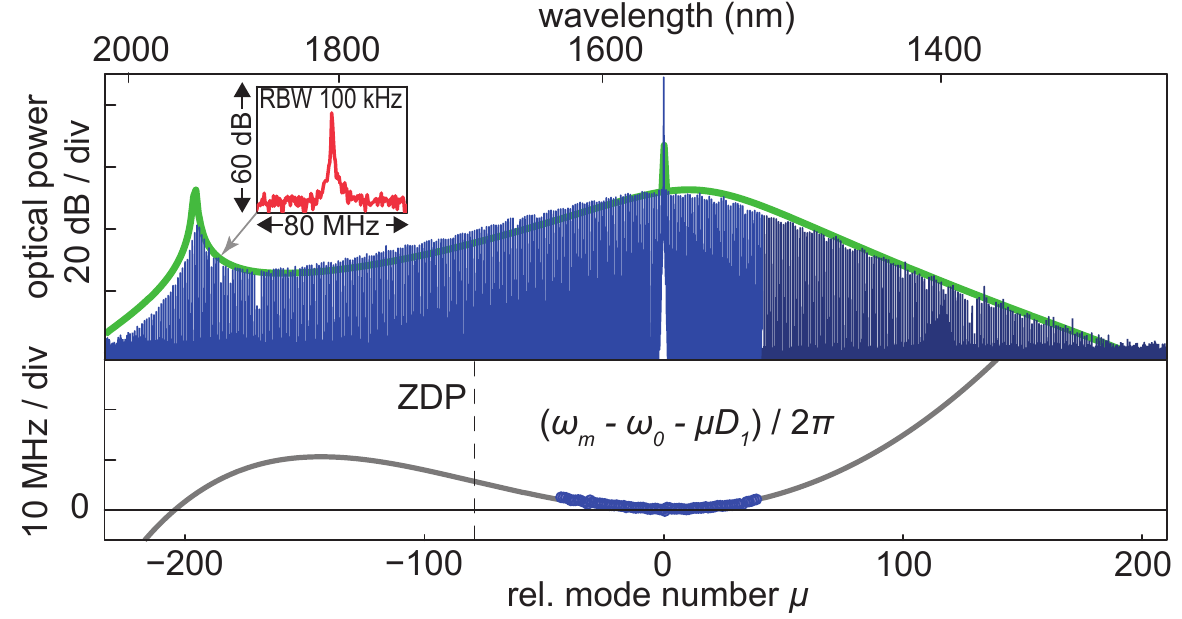}

\protect\protect\protect\protect\caption{\label{fig:Single-temporal-soliton-SiN}Single Kerr soliton generation
in a $\mathrm{Si_{3}N_{4}}$ microresonator. The upper panel shows
the single soliton spectrum that covers a spectral bandwidth of 2/3
of an octave. The black line in the background is the spectral envelope
obtained by the simulation described in Section \ref{sec:Simulating Soliton Formation}.
The lower panel shows the measured dispersion (dots) free of avoided
mode crossing and the dispersion over the full spectral span as obtained
through finite element simulation. The dispersive wave forms at the
wavelength of approximately 1.9 micron where the phase matching condition
$\omega_{\mu}-\omega_{0}-\mu\cdot D_{1}\approx0$ is fulfilled. The
zero dispersion point (ZDP) is marked by a vertical, dashed line.
The inset in the upper panel show a heterodyne beatnote between the
dispersive wave and an external laser. Its narrow width proves the
coherence of the dispersive wave.}
\end{figure}

The discovery of temporal DKS in crystalline resonators, as described
in the previous chapters not only enables generating coherent optical
frequency combs in a reliable manner, but also opened the ability
to use time domain broadening methods to achieve a coherent link from
the RF (or microwave) domain to the optical frequency domain. A key
emerging research question has been if solitons can also be generated
in other microresonator platforms. One particularly promising platform
for applications are $\mathrm{Si_{3}N_{4}}$ integrated microresonators.
This platform allows for the integration of coupling waveguide and
resonators on a microchip. Indeed, it was demonstrated that by carefully
optimizing the fabrication process, the influence of detrimental avoided
mode crossings can be strongly reduced, enabling the generation of
DKS in $\mathrm{Si_{3}N_{4}}$ microresonators\cite{Brasch2014b}.
Here, as in the case of crystalline resonators, the soliton formation
is indicated by discrete steps in the resonator transmission. Similar
to the method described above in detail in Section \ref{sec:Laser Tuning into Solitons}
a suitable laser scan allows for stable and thermally locked solitons
in $\mathrm{Si_{3}N_{4}}$ microresonators. Figure \ref{fig:Single-temporal-soliton-SiN}
shows the generation of a single soliton state in a $\mathrm{Si_{3}N_{4}}$
microresonator. The bandwidth of the generated spectrum is much wider
than what has been observed in crystalline resonators. As an immidiate
consequence higher order dispersion terms become relevant; indeed,
the third order dispersion causes the Kerr soliton to emit what
is known as \textit{Soliton-Cherenkov} radiation\cite{Akhmediev1995}
(or dispersive wave), i.e. a radiative tail in the time domain. It
is important to note that the process can also be understood in the
frequency domain as successive FWM \cite{Erkintalo2012}. Dispersive
wave formation has been predicted particularly for $\mathrm{Si_{3}N_{4}}$
resonators, where for a coherent spectrum a spectrally sharp dispersive
wave is expected\cite{Coen2013}, as it is indeed the case (cf. Figure
\ref{fig:Single-temporal-soliton-SiN}). For a resonator with third
order dispersion (i.e. non-zero $D_{3}$), this occurs approximately
near the mode number $\mu\approx-3\frac{D_{2}}{D_{3}}$ (three times
the distance of the zero dispersion point from the pump wavelength,
$\mu\approx-\frac{D_{2}}{D_{3}}$) \cite{Brasch2014b}.

It is important to note that unlike the case of a soliton propagating
along an optical fibers where the dispersive wave is generally incoherent
with the soliton (i.e. propagating at a velocity different from the
soliton's velocity), the coherence can be maintained in case of the
microresonator owing to the periodic boundary conditions. This is
similar to the coherence of the dispersive wave in supercontinua that
are generated by a periodic train of ultra-short pulses. At the same
time, the presence of a dispersive wave can also occur in the case
of a high noise, non soliton state. The presence of a dispersive wave
therefore is not a proof of a coherent Kerr soliton state\cite{Erkintalo2014}.
Note, that the coherence of the spectrum in Figure \ref{fig:Single-temporal-soliton-SiN}
has been verified by recording the soliton pulse repetition rate beatnote
at 190 GHz as well as the heterodyne beatnotes of the soliton spectrum
with external CW lasers at various wavelengths (including the dispersive
wave spectral region). Based on the spectral width a soliton pulse
duration of approximately 30 fs can be estimated.

\begin{figure}[H]
	\includegraphics[width=1\textwidth]{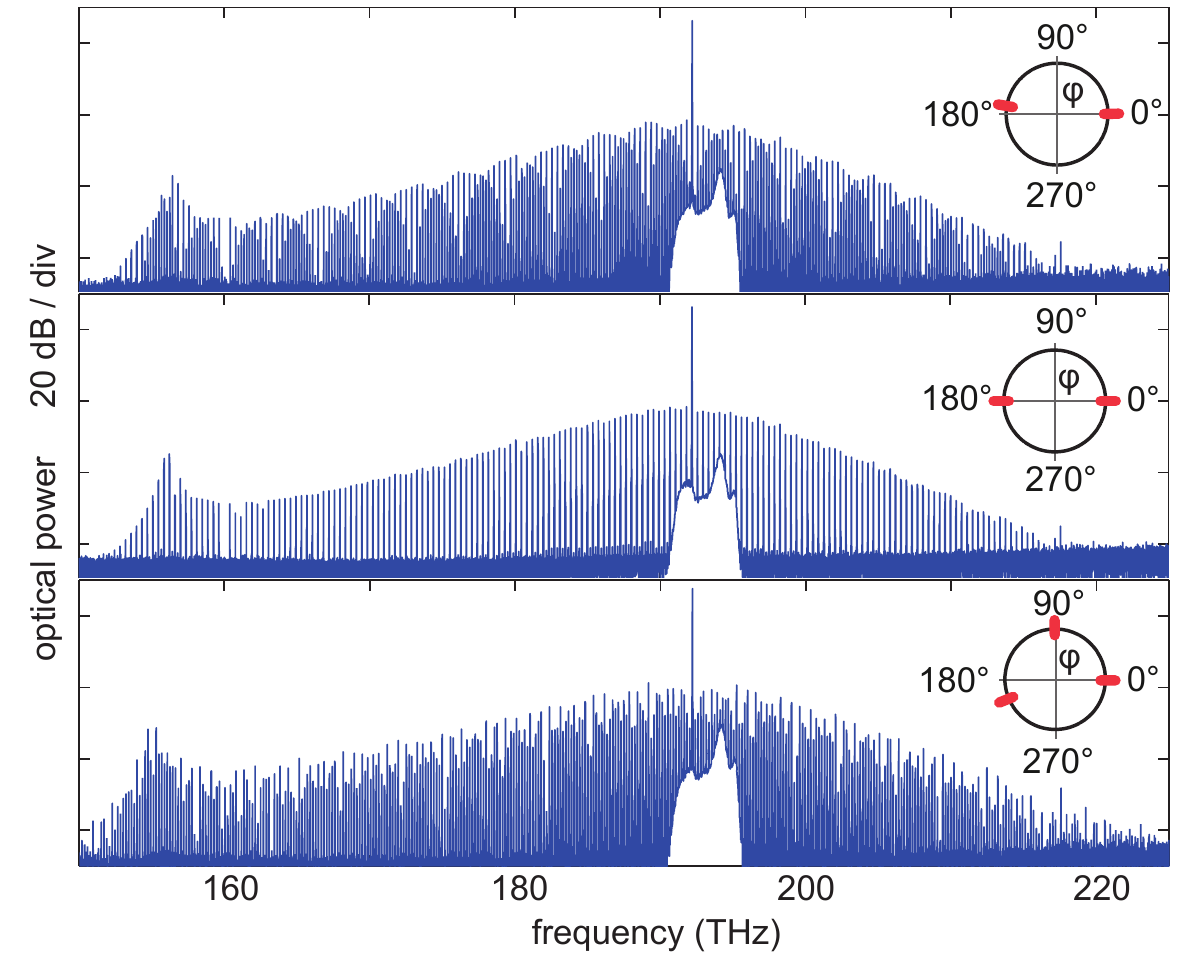}
	
	\protect\protect\protect\protect\caption{\label{fig:Generation-of-multiple-Solitons SiN}Generation of multi-soliton
		states and soliton induced Cherenkov radiation in a $\mathrm{Si_{3}N_{4}}$
		microresonator. The characteristic spectral modulations occur due
		to the interference of solitons at different positions in the microresonators
		as indicated in the insets.}
\end{figure}

It is also possible to generate multiple solitons in the $\mathrm{Si_{3}N_{4}}$
microresonators. As in the case of the crystalline resonators, once
a multi-soliton state is generated, the modulation of the spectral
envelope, resulting from the interference of the solitons, remains
stable for hours. The spectral envelope function $I(\mu)$ of this
interference is given by 
\begin{eqnarray*}
I(\mu)=\left|\sum\limits _{j=1}^{N}\mathrm{exp}(i\phi_{j}\mu)\right|^{2},
\end{eqnarray*}
where $\phi_{j}$ corresponds to the relative angular position of
the $j^{\mathrm{th}}$ soliton. From the envelope of the spectrum
the single soliton spectrum and the relative positions of the solitons
can be reconstructed as shown in Figure \ref{fig:Generation-of-multiple-Solitons SiN}.

\section{Summary}

\label{Sect:Model}

The observation and stable generation of temporal DKS
in optical microresonators provides a major impetus to the field of
microresonator frequency combs. It provides a long sought method for
the generation of widely spaced frequency combs, with smooth spectral
envelope and with low phase noise. In addition, soliton induced Cherenkov
radiation provides a method to increase the bandwidth and transfer
the coherence of the comb into the normal GVD regime. Together with
the ability to accurately predict the comb performance via numerical
simulations, this implies that the synthesis of octave spanning frequency
comb spectra directly from a CW laser using a microresonator is a
possibility. Solitons in microresonators may make frequency comb metrology
ubiquitous by enabling compact comb generators that exhibit mode spacings
in the $10-100$ $\mathrm{GHz}$ regime. Potential applications include
frequency metrology (i.e. frequency measurements), data transfer,
microwave signal generation, spectroscopy, optical sampling and arbitrary
optical waveform generation.





\bibliographystyle{unsrt}


\end{document}